\documentclass[10pt,a4paper,twoside]{article}

\usepackage{umlaute,eurosym,comment,lscape}
\usepackage{amsmath,amssymb,mathrsfs,amsbsy,bm,upgreek,theorem}
\usepackage{float,epsfig,subfig,wrapfig,xcolor,colortbl}
\usepackage{pstricks,pst-node}
\usepackage{longtable,multirow}
\usepackage[absolute,overlay]{textpos}
\usepackage{graphicx}
\usepackage{color}
\usepackage{makecell}
\graphicspath{{figures/}}
\usepackage{refcount}
\newcounter{auxcount}

\newcounter{FI}  % functional insufficiencies
\providecommand{\FI}{\refstepcounter{FI}FI\ifnum\value{FI}<10 0\fi\ifnum\value{FI}<100 0\fi\arabic{FI}}
\providecommand{\FIref}[1]{\setcounterref{auxcount}{#1}FI\ifnum\value{auxcount}<10 0\fi\ifnum\value{FI}<100 0\fi\ref{#1}}

\newcounter{TC}  % triggering conditions
\providecommand{\TC}{\refstepcounter{TC}TC\ifnum\value{TC}<10 0\fi\ifnum\value{TC}<100 0\fi\arabic{TC}}
\providecommand{\TCref}[1]{\setcounterref{auxcount}{#1}TC\ifnum\value{auxcount}<10 0\fi\ifnum\value{TC}<100 0\fi\ref{#1}}

\newcounter{Spec}  % specifications
\providecommand{\Spec}{\refstepcounter{Spec}S\ifnum\value{Spec}<10 0\fi\arabic{Spec}}
\providecommand{\Specref}[1]{\setcounterref{auxcount}{#1}S\ifnum\value{auxcount}<10 0\fi\ref{#1}}

\newcounter{HZ}  % hazards
\providecommand{\HZ}{\refstepcounter{HZ}Haz\ifnum\value{HZ}<10 0\fi\arabic{HZ}}
\providecommand{\HZref}[1]{\setcounterref{auxcount}{#1}Haz\ifnum\value{auxcount}<10 0\fi\ref{#1}}

\newcounter{SG}  % safety goals
\providecommand{\SG}{\refstepcounter{SG}SG\ifnum\value{SG}<10 0\fi\arabic{SG}}
\providecommand{\SGref}[1]{\setcounterref{auxcount}{#1}SG\ifnum\value{auxcount}<10 0\fi\ref{#1}}

\newcounter{SR}  % safety requirements
\providecommand{\SR}{\refstepcounter{SR}Req\ifnum\value{SR}<10 0\fi\arabic{SR}}
\providecommand{\SRref}[1]{\setcounterref{auxcount}{#1}Req\ifnum\value{auxcount}<10 0\fi\ref{#1}}

\newcounter{IR}  % integration requirements
\providecommand{\IR}{\refstepcounter{IR}IR\ifnum\value{IR}<10 0\fi\arabic{IR}}
\providecommand{\IRref}[1]{\setcounterref{auxcount}{#1}IR\ifnum\value{auxcount}<10 0\fi\ref{#1}}

\psset{unit=1mm,framesep=10pt,fillstyle=none}  % pstricks basic settings
\degrees[360]                                  % segmentation of the unit circle
\psset{linewidth=0.9pt}                        % standard line width
\definecolor{ULOcean}{RGB}{0,75,90}
\definecolor{ULOrange}{RGB}{236,116,4}
\definecolor{LiteRed}{RGB}{240,170,170}
\definecolor{MedRed}{RGB}{204,24,24}
\definecolor{DarkRed}{RGB}{127,15,15}
\definecolor{LiteYellow}{RGB}{255,245,128}
\definecolor{MedYellow}{RGB}{255,237,32}
\definecolor{DarkYellow}{RGB}{255,235,0}
\definecolor{LiteBlue}{RGB}{170,184,255}
\definecolor{MedBlue}{RGB}{43,78,255}
\definecolor{DarkBlue}{RGB}{0,42,120}
\definecolor{LiteGreen}{RGB}{207,224,207}
\definecolor{MedGreen}{RGB}{40,120,40}
\definecolor{DarkGreen}{RGB}{0,80,0}
\definecolor{LiteViolet}{RGB}{207,195,220}
\definecolor{MedViolet}{RGB}{80,29,127}
\definecolor{DarkViolet}{RGB}{53,0,107}
\definecolor{LiteOrange}{RGB}{255,220,187}
\definecolor{MedOrange}{RGB}{255,150,50}
\definecolor{DarkOrange}{RGB}{255,124,0}
\definecolor{LiteBrown}{RGB}{225,192,180}
\definecolor{MedBrown}{RGB}{154,40,0}
\definecolor{DarkBrown}{RGB}{116,30,0}
\definecolor{LiteGray}{RGB}{230,230,230}
\definecolor{MedGray}{RGB}{160,160,160}
\definecolor{DarkGray}{RGB}{100,100,100}

\definecolor{opred}{RGB}{255,150,150}
\definecolor{opgray}{RGB}{200,200,200}

\newlength\figureheight	
\newlength\figurewidth

\usepackage{lineno,hyperref}
\modulolinenumbers[5]
\begin{document}

\title{Analysis of Functional Insufficiencies and Triggering Conditions to Improve the SOTIF of an MPC-based Trajectory Planner}
%\tnoteref{mytitlenote}}
%\tnotetext[mytitlenote]{Footnote text goes here.}

\renewcommand{\thefootnote}{\fnsymbol{footnote}}
% Authors
\author{Mirko Conrad\footnote{samoconsult GmbH, Franz{\"o}sische Str.\ 13, 10117 Berlin, Germany}, Georg Schildbach\footnote{University of L{\"u}beck, Ratzeburger Allee 160, 23562 L{\"u}beck, Germany\newline\hspace*{0.5cm} corresponding author (\texttt{georg.schildbach@uni-luebeck.de})}}
\renewcommand{\thefootnote}{\arabic{footnote}}
\setcounter{footnote}{0}

%\cortext[mycorrespondingauthor]{}
%\ead{}

\maketitle

\section*{Abstract}
Automated and autonomous driving has made a significant technological leap over the past decade. In this process, the complexity of algorithms used for vehicle control has grown significantly. Model Predictive Control (MPC) is a prominent example, which has gained enormous popularity and is now widely used for vehicle motion planning and control. 
However, safety concerns constrain its practical application, especially since traditional procedures of functional safety (FS), with its universal standard ISO\,26262, reach their limits. Concomitantly, the new aspect of safety-of-the-intended-function (SOTIF) has moved into the center of attention, whose standard, ISO\,21448, has only been released in 2022.  Thus, experience with SOTIF is low and few case studies are available in industry and research. 
Hence this paper aims to make two main contributions: (1) an analysis of the SOTIF for a generic MPC-based trajectory planner and (2) an interpretation and concrete application of the generic procedures described in ISO\,21448 for determining functional insufficiencies (FIs) and triggering conditions (TCs). Particular novelties of the paper include an approach for the out-of-context development of SOTIF-related elements (SOTIF-EooC), a compilation of important FIs and TCs for a MPC-based trajectory planner, and an optimized safety concept based on the identified FIs and TCs for the MPC-based trajectory planner.

\section*{Keywords}
Automated Driving System (ADS)\;,\;\;
Autonomous Vehicles\;,\;\;
Trajectory Planning\;,\;\;
Model Predictive Control (MPC)\;,\;\;
Functional Safety (FS)\;,\;\;
Safety of the Intended Functionality (SOTIF)\;,\;\;
Functional Insufficiencies (FIs)\;,\;\;
Triggering Conditions (TCs)\;,\;\;
SOTIF-related Element out of Context (SOTIF-EooC)

\section{Introduction}

Since the beginning of the 21st century, automated and autonomous vehicles have become a major subject of research across the globe, in academia and industry \cite{MonteEtAl:2006,MonteEtAl:2008,BachaEtAl:2008}. The technological progress is fueled by ever more powerful hardware for computation, communication, and sensing. Another relevant factor is the rapid advances in mathematical algorithms, especially in Artificial Intelligence (AI), Machine Learning (ML), and Numerical Optimization (NO). A plethora of new and powerful methods have been developed and successfully demonstrated in actual vehicles on the road, among them \emph{Model Predictive Control (MPC)}. 

MPC is a popular computer-based control method \cite{PadenEtAl:2016,DixitEtAl:2020} that has been frequently tested in automated driving system (ADS) prototypes \cite{MusaEtAl:2021}. Thus far, however, it has not been considered a viable option for series production. The main reason is its relatively high complexity, causing relatively high efforts in the development process and increasing the computational hardware requirements. Arguably, an even bigger challenge results from safety concerns related to MPC. These concerns are similar to other state-of-the-art methods of AI, ML, and NO. 

Due to the complexity of these methods, and the complexity of the corresponding automated driving tasks, \emph{Functional Safety (FS)}, which is concerned with safety in the presence of malfunctions, is no longer the single most important aspect of safety. Instead, \emph{Safety of the Intended Function (SOTIF)}, which is concerned with the  safety of the nominal function even if all system elements work as intended, becomes equally important. The intention of this paper is hence to present a contribution towards a viable SOTIF concept for an MPC-based motion planning module in automated driving systems (ADSs).

\subsection{Model Predictive Control}

Over the past decades, MPC has developed into a mainstream approach for the control of complex and safety-critical systems, including ADSs. The basic idea is to formulate and solve a finite-horizon optimal control problem (FHOCP) as a numerical optimization program in real time \cite{Mayne:2000,GruenePannek:2011,BoBeMo:2017,RaMaDi:2018}. The FHOCP contains a model of the system and naturally integrates constraints on the model states and control inputs. It can be used for \emph{motion planning} \cite{GoetteEtAl:2016,FranzeLucia:2016} and \emph{trajectory tracking control} \cite{FraschEtAl:2013,KongEtAl:2015,GutjahrEtAl:2017}, where feedback is introduced by a receding-horizon implementation. It can also be used as a hybrid module, covering some or all aspects of planning and tracking control \cite{Carvalho:2016}. It has even been used to cover some aspects of tactical planning, such as lane change decisions \cite{Schildi:2015,CesariEtAl:2017}.

The current trend is to use MPC as a motion planner, such as in the commercial product ProDriver by Embotech \cite{ProDriver}. The goal is to integrate as many aspects as possible into the motion planning module, in order to slim down the module interfaces. Interfaces always require additional engineering efforts and they are a common source of functional, and thus also safety problems. The MPC module should thus be capable of making tactical decisions, by introducing integer variables, and ideally should also provide a basic level of vehicle and actuator control.

Recent research in the context of MPC for ADSs has begun to focus on interaction-aware motion planning \cite{CarvalhoEtAl:2015}. The basic idea is to consider the predictions of dynamic objects in the environment not as invariably fixed, but as interdependent with the other objects and also dependent on the actions of the ego vehicle. Hence the future trajectories of dynamic objects are included in the model-based predictions in the MPC. The interactions themselves can be modeled by different approaches, such as game theory \cite{LiuEtAl:2022} or neural networks \cite{GuptaEtAl:2023}. 

A common idea is to represent the intention of other drivers as modes \cite{NairEtAl:2022,BencioliniEtAl:2023}, corresponding to lane keeping or a lane change, for example. The remaining uncertainty regarding the exact trajectory of the dynamic object can be considered as stochastic noise. Another noteworthy approach is Branch MPC \cite{ChenEtAl:2022,OliveiraEtAl:2023}, where the modes of each dynamic object and the ego vehicle may change at several specific instances over the prediction horizon.

\subsection{Safety of the Intended Function (SOTIF)}

SOTIF is a comparatively new branch of safety that is becoming increasingly important for SAE driving automation levels 3 or higher. The first edition of the applicable SOTIF standard ISO\,21448 has only been published in 2022 \cite{ISO21448:2022}. 
Due to its relatively recent introduction, scientific literature on the application of SOTIF to ADSs is still  sparse. % R. Zhu, A. Gu, Z. Wu, B. Liu and M. Yu, "Research on SOTIF of automatic driving system," 2022 14th International Conference on Measuring Technology and Mechatronics Automation (ICMTMA), 2022, pp. 228-231, doi: 10.1109/ICMTMA54903.2022.00051.
Zhu et.\ al.\ \cite{9724010} emphasize that the research on the SOTIF of automated and autonomous driving systems is still in its infancy. Their paper focuses on safety analysis methods that could be applied to SOTIF, but it does not cover specifically the topic of identifying and analyzing functional insufficiencies (FIs) or triggering conditions (TCs).

% Z. Qidong et al., "The Research on the Identification of ACC SOTIF Triggering Conditions Based on Scenario Analysis," 2022 IEEE International Conference on Real-time Computing and Robotics (RCAR), 2022, pp. 263-266, doi: 10.1109/RCAR54675.2022.9872207.
Qidong et.\ al.\ \cite{9872207} state that the identification and systematic analysis of FIs and TCs remains a major challenge for the SOTIF. Using an Adaptive Cruise Control (ACC) system as an example, the authors propose a scenario-based analysis method of SOTIF triggering conditions. In this approach, the TCs are analyzed according to elements of SOTIF scenarios that are formed in view of FIs. 
The approach is then applied to the 'Sense' portion of the ACC system. The 'Plan' part is not addressed in this paper.
% "If the system has no fault, it can be assumed that relevant calculation results are accurate and effective. Accordingly, possible functional insufficiencies are mostly caused by late or inaccurate identification of the target vehicle".

% L. Junfeng, Z. Yunshuang, Z. Shuai, C. Chao and D. Zhibin, "A Research on SOTIF of LKA based on STPA," 2022 IEEE International Conference on Real-time Computing and Robotics (RCAR), 2022, pp. 396-400, doi: 10.1109/RCAR54675.2022.9872242.
Junfeng et.\ al.\ \cite{9872242} apply the System-Theoretic Process Analysis (SPTA) method to study the SOTIF of a Lane Keep Assistance (LKA) system. The study intends to verify the feasibility of STPA as part of the SOTIF lifecycle.
The paper proposes a set of TCs for the LKA, but not all of them seem to be TCs in the sense of ISO\,21448. A subset of these TCs will be used as input for the SOTIF analysis of the MPC-TP. 
In the analysis, there is also no specific focus on the 'Plan' part.

\subsection{Content and Contributions}

This paper describes a possible approach for the analysis of FIs and TCs, with the goal of improving the SOTIF of a generic planning module for an ADS. The trajectory planning module is based on the Model Predictive Control (MPC) approach. It is assumed that the MPC-based trajectory planner (MPC-TP) can be integrated into different vehicle-level functions and that it is re-usable for different vehicles within a specified vehicle class. Hence it is treated as a \emph{SOTIF-related element out of Context (SOTIF-EooC)}, which is a particular novelty compared to the existing literature.

The main objective of this paper is to demonstrate how the framework of ISO\,21448 \cite{ISO21448:2022} can be used for improving the SOTIF of a modern, state-of-the-art planning algorithm. The system architecture and functional specifications are outlined in Section \ref{Sec:Specification} and the details of the MPC-TP are provided in Section \ref{Sec:DevelopmentScope}. The relevant SOTIF background is outlined in Section \ref{Sec:SOTIFBackground}. The main part of this paper is the SOTIF analysis in Section \ref{Sec:SOTIFAnalysis}, which describes the application of various methods and techniques from ISO\,21448 \cite{ISO21448:2022} towards the identification of FIs and TCs for the MPC-TP. The analysis leads to a refined specification and design of the MPC-TP, as described in Section \ref{Sec:RefinedSpecDesign}. The conclusions of this work are presented in Section \ref{Sec:Conclusion}. The main contributions of this paper can be summarized as follows:
\begin{itemize}
\item Application of the abstract / generic framework of the new SOTIF standard ISO\,21448 to a realistic case study, with a focus on early parts of the SOTIF lifecycle;
\item Interpretation and application of generic methods in ISO\,21448 for the determination of FIs / TCs, in the context of a specific use case;
\item Derivation of a structured list of FIs / TCs as a consolidated result of the SOTIF analysis (in a tabular format);
\item A refinement of the notion of an out-of-context development in the scope of SOTIF.
\end{itemize}

%The project Embedded Excellence \textcolor{red}{for Vehicle Motion} (EEmotion), funded by the German Federal Ministry for Economic Affairs and Climate Action, intends to integrate these algorithms into the planning and control modules of an automated driving system (ADS). The goal is to make the actions taken by a vehicle more intelligent, in order to improve the energy efficiency or to increase the individual ride quality. 

%A critical aspect for the realization of an ADS is \emph{safety}, in particular if learning-based or optimal \textcolor{red}{[optimization?] }algorithms are integrated. Different aspects of safety have to considered in the development of an ADS, including \emph{functional safety}, and \emph{safety of the intended functionality (SOTIF)}. 

%\section{Specification of the Functionality and Considerations for the Design of the Vehicle Level Control System}\label{Sec:Specification}

\section{Initial Concept for Vehicle Level Control System}\label{Sec:Specification}

% Starting point for the SOTIF activities is the \emph{Specification and Design} of the item under  consideration \cite[Section 5]{ISO21448:2022}. It may comprise requirements, functional and design specifications, and the system architecture. In fact, the initial specification and design are iterated and refined over multiple cycles until the overall SOTIF risk is deemed acceptable. During these cycles, the goal is to resolve or improve all relevant functional insufficiencies.

%In our case study, the Specification and Design mainly includes the system architecture (Section \ref{Sec:Architecture}) and the functional specifications at the vehicle level (Section \ref{Sec:Function}), as well as the design specification of the MPC-TP at the element level (Section \ref{Sec:MPC-TP}).

The starting point for the design of the planning module as a SOTIF-SEooC are assumptions about the encompassing vehicle control system into which the module is to be embedded, as well as the related interfaces. Furthermore, specifications are needed about the overall functionality and the operational design domain (ODD). 

\subsection{Functional Architecture}\label{Sec:Architecture}

\begin{figure}[H]
	\includegraphics[width=1.0\textwidth]{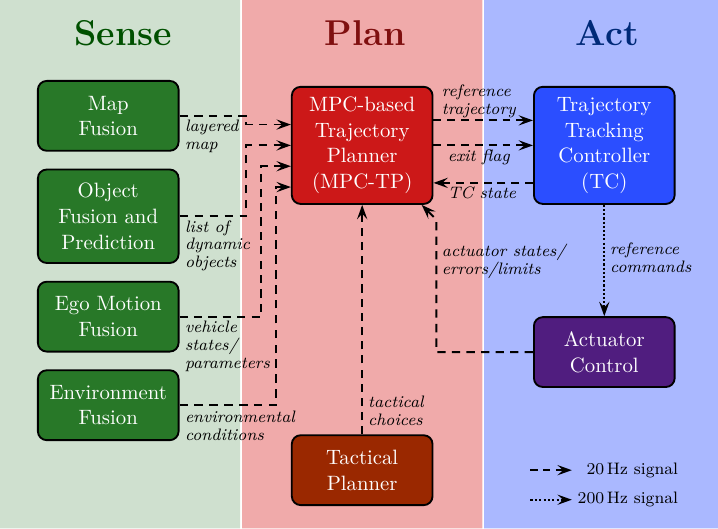}
	\caption{Overview of system architecture.\label{Fig:Architecture}}
\end{figure}

The functional architecture of a generic ADS is  shown in Figure \ref{Fig:Architecture}. The MPC-TP is shown in the middle (red). The task of the MPC-TP is to compute a reference trajectory, which is then used by the trajectory controller (blue) and lower-level modules to control the vehicle's actuators. The actuators are an integral part of the chassis, including, in particular, the drivetrain and braking systems for longitudinal motion control and the steering system for lateral motion control. 

The MPC-TP receives information from centralized estimation modules (green), which may serve also other functions and/or other modules in the architecture \cite{BaerEtAl:2009}. The key point is that they are developed independently of the MPC-TP. These modules provide a local map, including the topology and road geometry and information about the infrastructure, such as traffic signs (Map Fusion). They also include the relative position of the vehicle on the map, the vehicle's dynamic states, a list of relevant vehicle parameters, and diagnostic information about the vehicle's sensors and actuators (Ego Motion Fusion). Furthermore, a list of relevant traffic objects in the vicinity of the vehicle is provided, including their classification, position, speed, and other states, as well as a situation analysis and prediction(s) about (possible) future evolution of the situation (Object Fusion and Prediction). The sensors also deliver important information about the environment, such as ambient temperature or road surface (Environment Fusion). Finally, the MPC-TP may receive high-level commands by the human driver or the ADS, such as the reference speed, which lane or road to use, or where to take a turn (Tactical Planner).

%\begin{figure}[H]
%    \vspace*{0.2cm}
%    \begin{center} 
%\includegraphics[width=1.0\textwidth]{figures/Architecture.PNG}
%\end{center}
%\vspace*{-0.4cm}
	%\caption{Overview of system architecture.\label{Fig:Architecture}}
%\end{figure}

% The Sense Plan Act (SPA) model, which is used in the SOTIF standard ISO\,21448 \cite{ISO21448:2022}, is overlayed in Figure \ref{Fig:Architecture}. It splits an ADS into the three main components 'Sense', 'Plan', and “Act”. For example, the Object Fusion and Prediction block is allocated to ``Sense'' category, the trajectory planning belongs to the category ``Plan'', and the actuator control is assigned to ``Act''.  

\subsection{Encompassing Vehicle Control System}\label{Sec:VehCtrlSys}

The MPC-TP is treated as a SOTIF-related element out of context (SOTIF-EooC), as detailed in Section \ref{Sec:SOTIF-EooC}. Hence it shall be possible to integrate the MPC-TP into different the SOTIF-related systems as well as different vehicle types. For this paper, it is assumed that the MPC-TP can be used in the context of the following systems for a given class of passenger cars:
\begin{itemize}
    \item[F1)] \emph{Energy Efficient Lane Centering and Adaptive Cruise Control (E\textsuperscript{2}LCACC)}: E\textsuperscript{2}LCACC is a SAE level 2 function that combines the classic Lane Centering (LC) and Adaptive Cruise Control (ACC). It performs the lateral and longitudinal control of the vehicle, in order to keep the vehicle on a reference path in its lane and to follow the preceding vehicle. E\textsuperscript{2}LCACC contains AI-based elements for energy efficient driving, in particular to determine the optimal speed profile. 
    \item[F2)] \emph{Active Lane Change Assist (ALCA)}: ALCA is SAE level 2 or 3 function designed to help drivers changing lanes safely by automatically steering the vehicle into the adjacent lane. ALCA operates in combination with E\textsuperscript{2}LCACC and also provides longitudinal and lateral control of the vehicle during overtaking. The driver can issue lane change requests to the left or the right, e.g., by operating the indicator lever. The ALCA then checks whether a safe lane change is possible within 10 seconds. If it is possible, it performs the lane change; if it is not possible, it keeps the vehicle in the current lane and continues with E\textsuperscript{2}LCACC.
    \item[F3)] \emph{Highway Pilot (HP)}: The HP is a SAE level 4 function that takes full control of the vehicle during highway driving. Only high level objectives are provided by the user, such as the destination, the desired route, the targeted arrival time, or the driving style. The driving itself, including all decisions, is performed autonomously by the HP.
\end{itemize}

\subsection{Operational Design Domain (ODD)}\label{Sec:ODD}

The vehicle-level functions F1, F2, F3 are only to be used within the conditions and limits of the ODD ({\Spec\label{Spec:ODD}}) specified in Table \ref{Tab:ODDDescription}. The ODD documents the specific conditions under which the ADS is designed to function \cite{ISO21448:2022}. Its description is based on the structure provided in DOT\,HS\,812\,623 \cite{ThornEtAl2018}.

\begin{table}[H]
\small
\renewcommand{\arraystretch}{1.5}
\begin{tabular}{| p{3.7cm} | p{6cm} | p{1.0cm} |}
%\label{ODDinitial}
\hline 
  \multicolumn{2}{|l|}{\textbf{Physical Infrastructure}} & \textbf{Identi- fier}\\
\hline\hline
  roadway types: & German divided freeway (Autobahn) & (\Specref{Spec:ODD}a)  \\
  \hline 
  roadway surfaces: & asphalt or concrete slabs & (\Specref{Spec:ODD}b) \\
  \hline 
  roadway geometry: &  straightways or curves with bank angles $\leq8^{\circ}$ (max.\ value permitted for German roads) & (\Specref{Spec:ODD}c) \\
  \hline \hline 
  \multicolumn{3}{|l|}{\textbf{Objects}} \\
\hline\hline
  roadway users - vehicles: & all vehicles drive forward only & (\Specref{Spec:ODD}d)  \\
  \hline 
  non-roadway users - obstacles/objects: & no pedestrians or bicycles & (\Specref{Spec:ODD}e)\\
  \hline \hline  
  \multicolumn{3}{|l|}{\textbf{Environmental Constraints}} \\
\hline \hline 
  weather - temperature: & ambient temperature within $[-20^{\circ}\mathrm{C}, +50^{\circ}\mathrm{C}]$ & (\Specref{Spec:ODD}f) \\
  \hline 
  weather - wind: & wind speed < $8\,\mathrm{bft}$ ($\approx 74\,\frac{\mathrm{km}}{\mathrm{h}}$) & (\Specref{Spec:ODD}g) \\
     \hline 
  weather-induced roadway conditions - friction value: & $\mu_{\min} = 0.15$\;,\;\;$\mu_{\max} = 1.2$ & (\Specref{Spec:ODD}h) \\
  \hline

\hline \hline 
  \multicolumn{3}{|l|}{\textbf{Operational Constraints}} \\
\hline\hline
  speed limits - min.\ and max.\ speed limit: & $v_{\min}=0\;\frac{\mathrm{km}}{\mathrm{h}}$\;,\;\;$v_{\max}=130\;\frac{\mathrm{km}}{\mathrm{h}}$ & (\Specref{Spec:ODD}i)\\
  \hline 
  
\end{tabular}
\renewcommand{\arraystretch}{1.0}
\normalsize
\caption{Initial ODD.\label{Tab:ODDDescription}}
\end{table}

\subsection{Assumptions about the Functionality and the Vehicle Class}\label{Sec:Function}

Since the MPC-TP is considered as a SOTIF-EooC, specific assumptions are required about its integration into the vehicle and the vehicle-level functions F1, F2, F3, as well as its interfaces to other modules. An overview of the upstream and downstream modules can be gathered from Figure \ref{Fig:Architecture}. The perception modules make use of appropriate sensors and communication infrastructure. The control modules are connected with the relevant vehicle actuators. It is beyond the scope of this paper to specify these modules here in detail. A list of the most relevant specifications for the ego vehicle (EgoV) and the module interfaces is provided below.
\setlength{\leftmargini}{1.5cm}
\begin{itemize}
    \item[\Specref{Spec:ODD}j)] The EgoV belongs to a specific class of passenger vehicles, consisting of a finite number of vehicle types.
    \item[\Specref{Spec:ODD}k)] All possible EgoV types are known at the time of the MPC-TP development process, including their physical design and specifications, chassis and suspension systems, and chassis control systems (Anti-lock Braking System, Electronic Stability Control, etc.).
    \item[\Specref{Spec:ODD}l)] Further parameters of EgoV (such as the vehicle mass, location of the center of gravity, moments of inertia, and tire parameters) are known to be within specified value ranges.
    \item[\Specref{Spec:ODD}m)] The loads inside the EgoV are within the limits specified by the respective vehicle type.
    \item[\Specref{Spec:ODD}n)] There are no external loads attached to the EgoV, such as a trailers, roof racks, or any type of load carriers. 
\end{itemize}

\section{Initial Concept of the MPC-TP}\label{Sec:DevelopmentScope}

\begin{figure}[H]
    \hspace*{1.8cm}
        \includegraphics[width=1.0\textwidth]{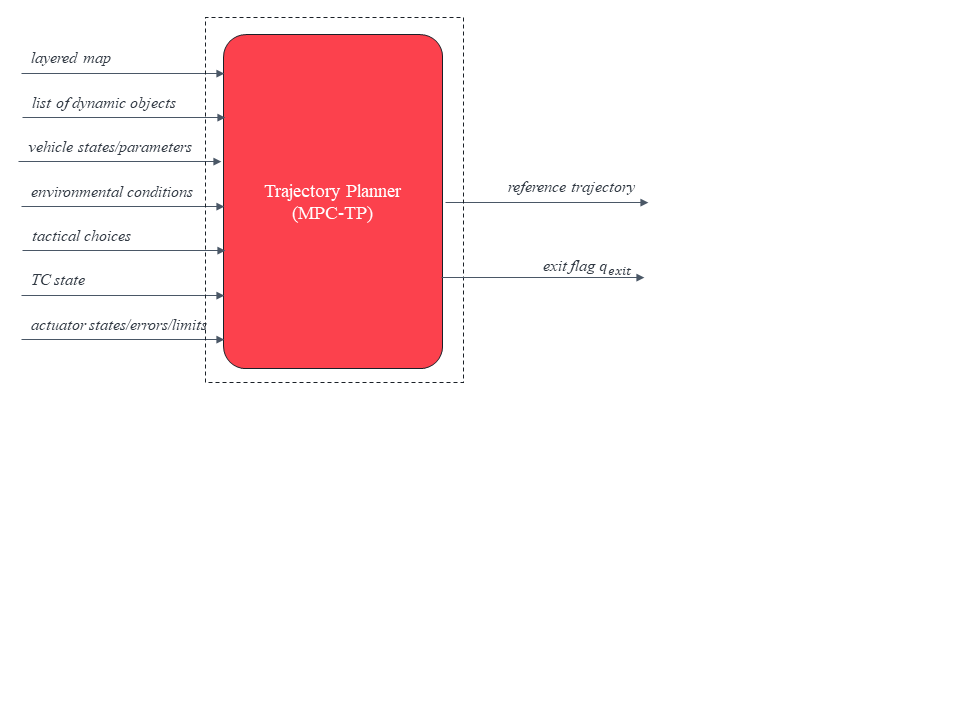}
    \vspace*{-4.8cm}
	\caption{Initial, single-block MPC-TP architecture.\label{Fig:MPC-TP_Initial}}
\end{figure}

The specifications for the MPC-TP are based on the overall architecture in Figure \ref{Fig:Architecture}. The MPC-TP is (initially) given by the single block displayed in Figure \ref{Fig:MPC-TP_Initial}. The specifications concern the input signals, the output signals, and the contents of the block itself, and they are presented along this structure in the remainder of this section.

\subsection{Input Interface / Perception}\label{Sec:MPCTPInterfaces}

\begin{figure}[H]
    \vspace*{0.2cm}
    \begin{center} 
\includegraphics[width=1.0\textwidth]{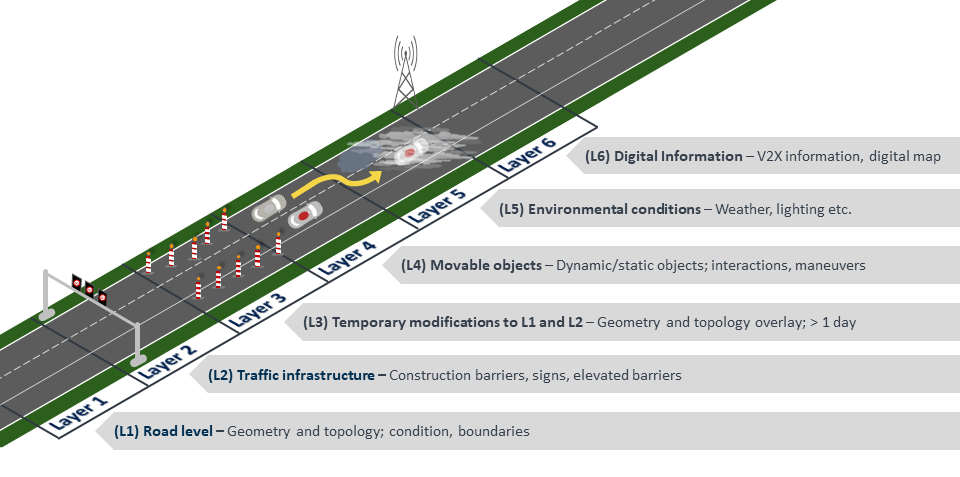}
\end{center}
\vspace*{-0.7cm}
	\caption{Illustration of the layered road model in ISO\,34503 \cite{PEGASUS:2019}.\label{Fig:LayeredRoadModel}}
\end{figure}

For the interface to the map fusion module, the \emph{6 layers road model} specified in ISO\,34503 \cite{ISO34503:2023} is considered. See Figure \ref{Fig:LayeredRoadModel} for an illustration.

\setlength{\leftmargini}{0.87cm}
\begin{itemize}
\item[{\Spec\label{Spec:MapFusion}})] \textit{Map Fusion:} The MPC-TP receives a layered map every $50\,\mathrm{ms}$, including
\begin{itemize}
    \item[\Specref{Spec:MapFusion}a)] geometry and topology of the road, lane and road boundaries (layer 1),
    \item[\Specref{Spec:MapFusion}b)] traffic signs, speed limits, traffic guidance, construction barriers (layer 2),
    \item[\Specref{Spec:MapFusion}c)] temporal modifications, e.g., due to construction sites (layer 3).
\end{itemize}
\end{itemize}
For the sake of the SOTIF-EooC, it is assumed that all of this map data is always fully accurate. The MPC-TP must plan its trajectories such that they conform to the applicable traffic rules (e.g., speed limit, no passing zone).

\begin{itemize}
\item[{\Spec\label{Spec:ObjectFusion}})] \textit{Object Fusion and Prediction:} The MPC-TP receives a complete list of dynamic objects in its relevant vicinity every $50\,\mathrm{ms}$, including
\begin{itemize}
    \item[\Specref{Spec:ObjectFusion}a)] other traffic participants (layer 4),
    \item[\Specref{Spec:ObjectFusion}b)] animals and other temporary obstacles, like dropped cargo, rocks, etc. (layer 4).
\end{itemize}
\end{itemize}
All dynamic objects include additional information about their classification (vehicle, truck, motorcycle, pedestrian, animal type, obstacle, etc.), position, velocity, and heading direction, if applicable. Furthermore, a reliable prediction for all of these objects is provided over the next 6 seconds. The uncertainty in these predictions is accounted for by the use of intervals (e.g., for the velocity) or bounding boxes (e.g., for the position).

No assumptions are made on specific sensors, their field-of-view, or their operational status. All assumptions are based on the interfaces to the sensor fusion modules. If the information flow on these interfaces cannot be guaranteed, the vehicle-level function will hand over to the driver (SAE levels 1,2,3) or transfer the vehicle to a safe state (SAE levels 4,5).

\begin{itemize}
\item[{\Spec\label{Spec:EgoFusion}})] \textit{Ego Motion Fusion:} The MPC-TP receives relevant information about the states and parameters of the ego vehicle every $50\,\mathrm{ms}$, along with a with maximum estimation error (marked with a ``$\Delta$''):
\begin{itemize}
    \item[\Specref{Spec:EgoFusion}a)] the relative position of the ego vehicle on the map $x,y$ $(\Delta x, \Delta y)$,
    \item[\Specref{Spec:EgoFusion}b)] the ego vehicle velocity $v_{\mathrm{lon}},v_{\mathrm{lat}}$ $(\Delta v_{\mathrm{lon}},\Delta v_{\mathrm{lat}})$,
    \item[\Specref{Spec:EgoFusion}c)] the ego vehicle's global heading angle $\psi$ $(\Delta\psi)$,
    \item[\Specref{Spec:EgoFusion}d)] the ego vehicle's acceleration $a_{\mathrm{lon}},a_{\mathrm{lat}}$ $(\Delta a_{\mathrm{lon}},\Delta a_{\mathrm{lat}})$,
    \item[\Specref{Spec:EgoFusion}e)] the vehicle's mass $m$ ($\Delta m$),
    \item[\Specref{Spec:EgoFusion}f)] the vehicle's moment of inertia around the vertical axis $I_{z}$ ($\Delta I_{z}$),
    \item[\Specref{Spec:EgoFusion}g)] the position of the center of gravity relative to the midpoint of the rear axle $x_{\mathrm{cog}}$, $y_{\mathrm{cog}}$, $z_{\mathrm{cog}}$ ($\Delta x_{\mathrm{cog}}$, $\Delta y_{\mathrm{cog}}$, $\Delta z_{\mathrm{cog}}$),
    \item[\Specref{Spec:EgoFusion}h)] the cornering stiffness of the front and the the rear tires $C_{\alpha,\mathrm{f}}$, $C_{\alpha,\mathrm{r}}$ ($\Delta C_{\alpha,\mathrm{f}}$, $\Delta C_{\alpha,\mathrm{r}}$).
\end{itemize}

\item[{\Spec\label{Spec:TrackControl}})] \textit{Tracking Controller:} Further information about the ego vehicle, in particular about the states of the lower-level actuators, are provided to the MPC-TP through the Tracking Controller every $50\,\mathrm{ms}$:
\begin{itemize}
    \item[\Specref{Spec:TrackControl}a)] the front wheels' steering angles $\delta_{\mathrm{f},\mathrm{l}}$, $\delta_{\mathrm{f},\mathrm{r}}$,
    \item[\Specref{Spec:TrackControl}b)] the current engine torque $T_{\mathrm{eng}}$,
    \item[\Specref{Spec:TrackControl}c)] the total gear ratio $i_{\mathrm{gear}}$,
    \item[\Specref{Spec:TrackControl}d)] the effective brake torque $T_{\mathrm{brake}}$.
    \end{itemize}

\noindent Actuator failures, controller failures or performance degradations are not considered in this paper, as the focus is on SOTIF instead of functional safety.

\item[{\Spec\label{Spec:EnvironmentFusion}})] \textit{Environment Fusion:} The MPC-TP receives accurate information about the environmental conditions every $50\,\mathrm{ms}$, including
\begin{itemize}
    \item[\Specref{Spec:EnvironmentFusion}a)] conditions of the road surface, in particular the friction coefficient $\mu$ ($\Delta\mu$) (layer 1),
    \item[\Specref{Spec:EnvironmentFusion}b)] weather conditions and lighting conditions, in particular the side wind speed $v_{\mathrm{wind}}$ ($\Delta v_{\mathrm{wind}}$) (layer 4).
\end{itemize}
\end{itemize}
In reality, environment parameters, such as the friction coefficient or the side wind speed, are variable over space and/or time. For simplicity, however, the architectures presumes the communication of constant values, which characterize the entire drivable area of the ego vehicle over the relevant prediction time, i.e., in this case 6 seconds. The tolerance thus includes the measurement uncertainty as well as the true temporal and/or spacial variations, if applicable.

\begin{itemize}
\item[{\Spec\label{Spec:TacticalPlanning}})] \textit{Tactical Planner:} The MPC-TP receives tactical choices from the Tactical Planner every $50\,\mathrm{ms}$, containing
\begin{itemize}
  \item[\Specref{Spec:TacticalPlanning}a)] the desired \emph{target lane} (e.g., the current lane, the left or right lane), 
  \item[\Specref{Spec:TacticalPlanning}b)] a desired \emph{reference speed} (e.g., because the vehicle is about to leave the highway on the next exit),
  \item[\Specref{Spec:TacticalPlanning}c)] optional information about the desired lateral position on the target lane (e.g., deviation from the lane center line),
  \item[\Specref{Spec:TacticalPlanning}d)] optional information about desired distances to other vehicles or objects (e.g., the desired gap to the leading or trailing vehicle).
\end{itemize}
\end{itemize}

All requests by the Tactical Planner are facultative. For example, if the Tactical Planner requests a lane change, it is the task of the MPC-TP to evaluate if this lane change is feasible and safe, given the dynamics of the vehicle and the traffic situation (dynamic and static obstacles, traffic rules, etc.). If it is not feasible or not safe, the MPC-TP resorts to a trajectory that keeps the EgoV on its current lane. Similarly, the desired distances in (\Specref{Spec:TacticalPlanning}c,d) are for performance purposes only, and not required for safety.

The tactical choices are generated by the active driving function. For example, for the E$^2$LCACC the unique command is always to keep the current lane, for the ALCA the lane change commands are issued by the driver, and for the HP the commands are issued by a dedicated software module.

\subsection{Output Interface / Reference Trajectory}\label{Sec:RefTrajectory}

The main output of the MPC-TP is a \emph{reference trajectory}, as shown in the architecture in Figure \ref{Fig:Architecture}. The reference trajectory consists of (i) a trajectory header and (ii) a trajectory body. The \emph{trajectory header} contains a sequential number, the generation time, a validity time interval, and a checksum.

\begin{figure}[t]
	\includegraphics[width=1.0\textwidth]{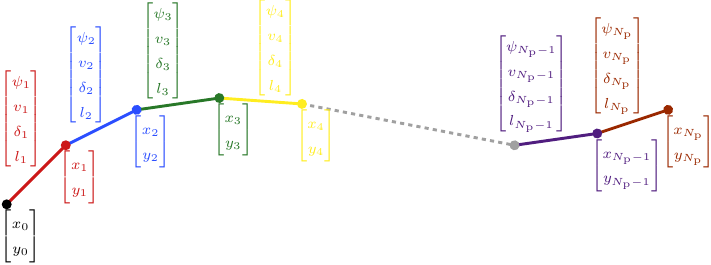}
	\vspace*{-0.7cm}
\caption{Illustration of the reference trajectory.\label{Fig:RefTrajectory}}
\end{figure}

\begin{itemize}
\item[{\Spec\label{Spec:TrajectoryBody}})] \textit{Trajectory Body:} The main part of the reference trajectory is the \emph{trajectory body}. It is represented by $N_{\mathrm{p}}-1$ linear segments, as illustrated in Figure \ref{Fig:RefTrajectory}. Each segment, numbered with $i=1,\dots,N_{\mathrm{p}}-1$, carries the following data:
\begin{itemize}
    \item[{\Specref{Spec:TrajectoryBody}}a)] $x_{i},y_{i}$: the planar coordinates of the starting point of the segment,
    \item[{\Specref{Spec:TrajectoryBody}}b)] $\psi_{i}$: the heading angle of the segment,
    \item[{\Specref{Spec:TrajectoryBody}}c)] $v_{i}$: the reference speed of the vehicle along the segment,
    \item[{\Specref{Spec:TrajectoryBody}}d)] $\delta_{i}$: the reference steering angle along the segment,
    \item[{\Specref{Spec:TrajectoryBody}}e)] $l_{i}$: the length of the segment.
    \item[{\Specref{Spec:TrajectoryBody}}f)] In addition, the final point of the trajectory is given in terms of its planar coordinates $x_{N_{\mathrm{p}}},y_{N_{\mathrm{p}}}$.
\end{itemize}
\end{itemize}
Note that the trajectory body data are partially redundant. This is mainly to avoid unnecessary and mirrored re-calculations within the MPC-TP.

\subsection{Specification and Design of the MPC-TP}\label{Sec:MPC-TP}

This case study considers a trajectory planner based on Model Predictive Control (MPC). A generic MPC formulation with a mainstream application to highway driving is used, in order to make the SOTIF concept in this paper as generally applicable as possible. It should be remarked, however, that numerous variations of MPC-based planning algorithms have been proposed in the recent literature \cite{GoetteEtAl:2016,LiuLeeEtAl:2017,YuEtAl:2021,MicheliEtAl:2022}. Moreover, commercial modules for trajectory planning with MPC have been made available, such as the software ProDriver by Embotech \cite{ProDriver}. It assumes a similar functional architecture as the MPC-TP in this paper, with a sampling time of $100\,\mathrm{ms}$. 

The computations of the MPC-TP are based on an internal prediction model. Following \cite{KongEtAl:2015}, a  \emph{dynamic bicycle model with a linear tire model} (LDBM) is selected and discretizes with an appropriate numerical integration method (e.g., Runge-Kutta):
\begin{equation}\label{Equ:LDBM}
\mathbf{x}_{t+1}=f(\mathbf{x}_{t},\mathbf{u}_{t})\;.
\end{equation}
The model is augmented with a suitable patch to circumvent the singularity of the tire forces at the origin. Here $t=0,1,2,\dots$ denote the discrete time steps, corresponding to multiples of the fundamental sampling time $t_{\mathrm{s}}=50\,\mathrm{ms}$. The vector $\mathbf{x}_{t}\in\mathbb{R}^6$ includes the \emph{(model) states} at step $t$, namely
\begin{itemize}
    \item $s_{t},e_{t}$: coordinates of the vehicle's reference point in road-aligned (Fren{\'e}t) coordinates, 
    \item $\psi_{t},\omega_{t}$: yaw angle and yaw rate of the vehicle,
    \item $v_{t},u_{t}$: longitudinal and lateral velocity of the vehicle.
\end{itemize}
The vector $\mathbf{u}_{t}\in\mathbb{R}^2$ contains the \emph{(control) inputs} at step $t$, namely
\begin{itemize}
    \item $\delta_{t}$: the steering angle of the front wheel,
    \item $M_{t}$: the torque of the drivetrain at the wheels, used to accelerate or decelerate the vehicle.
\end{itemize}
Furthermore, a variable for the \emph{reference lane} $n_{t}$ is used to indicate the lane decision for the vehicle at time $t$. It is assumed that the lanes are numbered from right to left, starting with 1 and counting up to the total number of lanes. The set of available lanes at step $t$ is denoted $\mathbb{N}_{t}$.

Based on the prediction, the concept of the MPC-TP is to set up and solve the following Optimal Control Problem (OCP) in each time step, using an appropriate numerical optimization solver:
\begin{subequations}\label{Equ:MSSP}\begin{align}
	\underset{n_{t},\mathbf{u}_{t},\mathbf{x}_{t}}{\mathrm{minimize}}\,\,\, &\ell_{\mathrm{s}}\bigl(s_{N_{\mathrm{p}}}\bigr)+\sum_{t=1}^{N_{\mathrm{p}}}\ell_{\mathrm{e}}\bigl(e_{t},n_{t}\bigr)+\ell_{\mathrm{o}}\bigl(s_{t},e_{t}\bigr)+\ell_{\mathrm{u}}\bigl(\delta_{t},M_{t}\bigr)\;,\\
	\text{s.t.}\,\,\,& \mathbf{x}_{t+1}=f\bigl(\mathbf{x}_{t},\mathbf{u}_{t}\bigr)\qquad \forall\;t=0,1,\dots,N_{\mathrm{p}}-1\;,\\
    &\mathbf{x}_{0}=\bar{\mathbf{x}}_{0}\;,\\
    &\mathbf{n}_{t}\in\mathbb{N}_{t}\qquad \forall\;t=1,2,\dots,N_{\mathrm{p}}\;,\\
	&\mathbf{u}_{t}\in\mathbb{U}\qquad \forall\;t=0,1,\dots,N_{\mathrm{p}}-1\;,\\
	&\mathbf{x}_{t}\in\mathbb{X}^{(\mathrm{s})}_{t}\;,\quad \mathbf{x}_{t}\in\mathbb{X}^{(\mathrm{t})}_{t}\qquad \forall\;t=1,2,\dots,N_{\mathrm{p}}-1\;,\\
    &\mathbf{x}_{N_{\mathrm{p}}}\in\mathbb{X}_{\mathrm{f}}\;.
\end{align}\end{subequations}
The objective function (\ref{Equ:MSSP}a) comprises the following cost terms:
\begin{itemize}
    \item $\ell_{\mathrm{s}}\bigl(s_{N_{\mathrm{p}}}\bigr)$: rewards the progress along the road made by the vehicle over the prediction horizon;
    \item $\ell_{\mathrm{e}}\bigl(e_{t},n_{t}\bigr)$: penalizes the lateral deviation of the vehicle from the center line of the corresponding reference lane;
    \item $\ell_{\mathrm{o}}\bigl(s_{t},e_{t}\bigr)$: repels the vehicle from static and dynamic obstacles, e.g., by means of a potential field;
    \item $\ell_{\mathrm{u}}\bigl(\delta_{t},M_{t}\bigr)$: favors a low input usage and smooth changes in the inputs, e.g., for driver comfort and energy efficient driving.
\end{itemize}
These cost terms are multiplied by individual tuning weights, which are selected during the control design process.

The initial state $\bar{\mathbf{x}}_{0}$ of the LDBM (\ref{Equ:MSSP}b,c) is obtained from the module for Ego Motion Fusion. The lane variables $n_{t}$ over the prediction horizon in (\ref{Equ:MSSP}d) are either a decision variable of the OCP (as for the HP), or they can be pre-determined before solving the OCP (as for E\textsuperscript{2}LCACC and ALCA). Equations (\ref{Equ:MSSP}e,f,g) represent the input constraints, state constraints, and terminal constraints, respectively. The state constraints are split into two sets:
\begin{itemize}
    \item The \emph{safety constraints} $\mathbb{X}^{(\mathrm{s})}_{t}$ avoid collisions of the EgoV with static and dynamic objects, most importantly other vehicles, called \emph{target vehicles} (TgtVs). Furthermore, the constraints ensure the validity of the tire model, by restricting the tire forces to remain within the admissible bounds of the linear tire regime (i.e., no slip in the tire-road contact patch).
    \item The \emph{traffic rule and comfort constraints} $\mathbb{X}^{(\mathrm{t})}_{t}$ represent the constraints from traffic rules (e.g., the EgoV must stay within its lane, the EgoV must respect the speed limit) and passenger comfort (e.g., no excessive acceleration or jerk).
\end{itemize}
The safety constraints are essential and cannot be relaxed, whereas the traffic rule and comfort constraints can be relaxed in emergency situations. Furthermore, in emergency situations, the LDBM in \eqref{Equ:LDBM} can be replaced with a nonlinear dynamic bicycle model (NDBM), which uses a Pacejka tire model instead of the linear tire model. By these modifications, the MPC-TP may temporarily operate the vehicle close to its dynamic limits and against the traffic rules, e.g., driving in between two lanes, or without considering passenger comfort.

\begin{itemize}
\item[{\Spec\label{Spec:MPC}})] Further specifications related to the MPC formulation include the following:
\begin{itemize}
    \item[\Specref{Spec:MPC}a)] The fundamental sampling time is $t_{\mathrm{s}}=50\,\mathrm{ms}$.
    \item[\Specref{Spec:MPC}b)] The planning horizon is $N_{\mathrm{p}}=120$ steps, which corresponds to a prediction time of $T_{\mathrm{p}}=6\,\mathrm{s}$.
    \item[\Specref{Spec:MPC}c)] The input constraint set $\mathbb{U}$ represents all limits of the steering, engine and braking systems, as well as the relevant limits of their rate-of-change.
    \item[\Specref{Spec:MPC}d)] The state constraint sets $\mathbb{X}_{t}$ are time-dependent and prevent collisions with static and dynamic obstacles, with the exception of vehicles driving behind the ego vehicle on the same lane. Moreover, the state constraints keep the vehicle away from its dynamic limits and ensure satisfaction of the traffic rules, e.g., by preventing forbidden lane changes or violation of the speed limit.
    \item[\Specref{Spec:MPC}e)] The terminal constraint $\mathbb{X}_{\mathrm{f}}$ corresponds to a safe state of the vehicle in which it can remain indefinitely (e.g., at standstill), or it represents a safe exit strategy (e.g., maximum deceleration until standstill). Its purpose is to ensure the recursive feasibility of the OCP.
\end{itemize}
\end{itemize}

\noindent The numerical computations of the MPC-TP comprise two main steps:
\begin{enumerate}
    \item \emph{setup} of the OCP problem,
    \item \emph{solution} of the OCP problem.
\end{enumerate}
Step 1 is to parse the OCP into the standard form of a nonlinear programming problem, which is realized by handwritten code. Step 2 is to solve the nonlinear program using a generic optimization solver, for which a generous selection of software packages is available \cite{Waechter:2010,Schildi:2016,GraichenEtAl:2019,ZanelliEtAl:2020}. 

\begin{itemize}
    \item[{\Spec\label{Spec:ExitFlag}})] Besides the optimal trajectory, the optimization solver returns an exit flag $q_{\mathrm{exit}}$ in exactly one of four classes: 
\begin{itemize}
    \setlength{\itemindent}{0.7cm}
    \item[$q_{\mathrm{exit}}=1$:] A (local) minimum is found with an acceptable tolerance level.
    \item[$q_{\mathrm{exit}}=2$:] A feasible point is found, but without an optimality property (e.g., because the gradient / search direction becomes too small or the solver has reached the maximum number of iterations / solution time).
    \item[$q_{\mathrm{exit}}=3$:] The solver detects that no feasible point exists (`certificate of infeasibility').
    \item[$q_{\mathrm{exit}}=4$:] A feasible point is not found, but without a certificate of infeasibility (e.g., because the solver has reached the maximum number of iterations / solution time), or an identified or unidentified solver error occurred (e.g., due to insufficient memory).
\end{itemize}
\end{itemize}

In reality, most solvers provide further diagnostics of the results, i.e., their exit flag is more detailed than the four classes above. The requirement (\Specref{Spec:ExitFlag}) states that it must be possible to map the true exit flag of the solver uniquely to one of the indicated classes. Moreover, it is assumed that the exit flag correctly characterizes the result of the solver. This can be verified by a separate solution checker. Cases where the exit flag is incorrect are not considered further as part of the SOTIF analysis. They represent software fault and hence belong to the realm of FS.

\subsection{Hazards at the Vehicle Level, Safety Goals, and SOTIF Requirements}\label{Sec:HazardsAndSOTIFReqs}

For the SOTIF analysis in this paper, the following \emph{hazards at the vehicle level} are considered:
\setlength{\leftmargini}{0.87cm}
\begin{itemize}
    \setlength{\itemindent}{0.4cm}
    \item[{\HZ\label{HZ:LeaveLane}})] The ego vehicle leaves the assigned lane.
    \item[{\HZ\label{HZ:Collision}})] The ego vehicle collides with other vehicles or obstacles.
\end{itemize}
These hazards lead directly to the specification of the following \emph{safety goals}:
\setlength{\leftmargini}{0.77cm}
\begin{itemize}
    \setlength{\itemindent}{0.4cm}
    \item[{\SG\label{SG:LeaveLane}})] The ego vehicle shall not leave the assigned lane(s).
    \item[{\SG\label{SG:Collision}})] The ego vehicle shall not collide with other vehicles or obstacles.
\end{itemize}

As part of a top-down safety lifecycle, these safety goals would be broken down recursively into more and more detailed safety requirements until one arrives at safety requirements for the component in question. Safety requirements for an ADS comprise both functional safety requirements and SOTIF-related safety requirements (or SOTIF  requirements in short).
Such a refinement of safety requirements does not take place when developing a SOTIF-EooC. Instead, safety requirements must be assumed for the out-of-context element under consideration. The SOTIF-SEooC's fulfillment of these \emph{assumed safety requirements} represents its contribution to achieving the original safety goals.
%\textcolor{red}{What is the connection to the safety goals?--> See above}
For the MPC-TP, we assume the following SOTIF-related safety requirements: 
\setlength{\leftmargini}{1.27cm}
\begin{itemize}
    \item[{\SR\label{SR:Timing}})] The MPC-TP shall provide a reference trajectory within $t_{\mathrm{c}}=30\,\mathrm{ms}$.\footnote{The computation time is less than the sampling time $t_{\mathrm{s}}=50\,\mathrm{ms}$, in order to account for additional communication delays.}
    \item[{\SR\label{SR:Consistent}})] The reference trajectory provided by the MPC-TP has to be \emph{consistent} regarding the redundant information in the trajectory body. 
    \item[{\SR\label{SR:Admissible}})] The reference trajectory provided by the MPC-TP has to be \emph{admissible}, in the sense that it must satisfy all relevant actuator constraints:
\begin{itemize}
    \item[\SRref{SR:Admissible}a)] The minimum / maximum longitudinal acceleration of the vehicle is $a_{\mathrm{lon},\min}=-10\,\frac{\mathrm{m}}{\mathrm{s}^2}$, $a_{\mathrm{lon},\max}=+3\,\frac{\mathrm{m}}{\mathrm{s}^2}$.
    \item[\SRref{SR:Admissible}b)] The minimum / maximum velocity of the vehicle is $v_{\min}=0\,\frac{\mathrm{km}}{\mathrm{h}}$, $v_{\max}=130\,\frac{\mathrm{km}}{\mathrm{h}}$.\footnote{The minimum velocity assumption means that the EgoV may not drive in reverse. A maximum velocity is required by the limited forward view of the vehicle's sensor suite.}
    \item[\SRref{SR:Admissible}c)] The maximum front wheel steering angle of the vehicle (in each direction) is $\delta_{\max}=34^{\circ}$.
    \item[\SRref{SR:Admissible}d)] The maximum front wheel steering rate (in each direction) is $\gamma_{\max}=68^{\circ}\,\frac{1}{\mathrm{s}}$.
    \item[\SRref{SR:Admissible}e)] The maximum lateral acceleration of the vehicle (in each direction) is $a_{\mathrm{lat},\max}=5\,\frac{\mathrm{m}}{\mathrm{s}^2}$.
\end{itemize}
\end{itemize}

Similar to the upstream perception modules, no detailed technical specifications have made for the downstream Tracking Controller (TC). As it is developed independently, responsibilities have to be clearly allocated and distributed between the MPC-TP and the TC. Thus requirements have to be specified for the interface, i.e., in this case, the reference trajectory. The approach proposed in this case study is to require ``drivability'' based on a nonlinear dynamic double-track model (NDDM):
\begin{itemize}
    \item[{\SR\label{SR:Drivable}})] The reference trajectory provided by the MPC-TP shall be \emph{drivable}, i.e., it must be possible to track it with a NDDM using some arbitrary controller, such that all of the following conditions are satisfied:
\begin{itemize}
    \item[\SRref{SR:Drivable}a)] The initial condition of the NDDM corresponds to the measured initial condition of the vehicle.
    \item[\SRref{SR:Drivable}b)] The parameters of the NDDM correspond to those of the vehicle (mass $m$, moment of inertia $I_{z}$, position of the center of gravity $x_{\mathrm{cog}}$ and $y_{\mathrm{cog}}$, front and rear cornering stiffness $C_{\alpha,\mathrm{f}}$ and $C_{\alpha,\mathrm{r}}$).
    \item[\SRref{SR:Drivable}c)] The tires of the NDDM do not saturate, i.e., there is no slip in the tire-road contact patch, for the worst-case friction coeffient $\mu-\Delta\mu$. This means that the tire forces remain inside the linear regime at all times.
    \item[\SRref{SR:Drivable}d)] The bank angle is between $\beta-\Delta\beta$ and $\beta+\Delta\beta$.
    \item[\SRref{SR:Drivable}e)] The side wind speed is between $v_{\mathrm{wind}}-\Delta v_{\mathrm{wind}}$ and $v_{\mathrm{wind}}+\Delta v_{\mathrm{wind}}$.
    \item[\SRref{SR:Drivable}f)] All existing actuator constraints are respected: communication delays, actuator dynamics, actuator limits, and limits of the rates of change.
    \item[\SRref{SR:Drivable}g)] The maximum lateral offset from the reference trajectory is $e_{\max}=5\,\mathrm{cm}$.
    \item[\SRref{SR:Drivable}h)] The length of each reference trajectory is such that it takes at least $5\,\mathrm{s}$ to reach the end.
\end{itemize}
\end{itemize}

The set of all \emph{consistent, admissible and drivable trajectories} is denoted as $\mathcal{T}$. Additionally, the trajectory should be such that it does not lead to collisions of the vehicle.
\begin{itemize}
\item[{\SR\label{SR:CollisionFree}})] The trajectory shall be \emph{collision-free}, i.e., it must stay within the boundaries of the road and it must avoid collisions with all dynamic obstacles. 
\begin{itemize}
    \item[\SRref{SR:CollisionFree}a)] For lateral collsition avoidance, the MPC-TP may rely on a maximum lateral offset $e_{\max}^{\mathrm{TC}}$ produced by the TC:
    \begin{itemize}
    \item for $v\leq 30\,\frac{\mathrm{km}}{\mathrm{h}}$: $e_{\max}^{\mathrm{TC}}=10\,\mathrm{cm}$,
    \item for $v\leq 80\,\frac{\mathrm{km}}{\mathrm{h}}$: $e_{\max}^{\mathrm{TC}}=15\,\mathrm{cm}$,
    \item for $v\leq 130\,\frac{\mathrm{km}}{\mathrm{h}}$: $e_{\max}^{\mathrm{TC}}=20\,\mathrm{cm}$.
    \end{itemize}
    \item[\SRref{SR:CollisionFree}b)] The MPC-TP shall plan its trajectories such that it is always safe for the ego vehicle to stay in its current lane, in particular by maintaining an appropriate safety distance to the leading vehicle.
\end{itemize}
\end{itemize} 
To this end, the MPC-TP may assume that all of the input information it receives is up-to-date and fully correct. 

Note that the scope of the paper is limited with regards to the considered vehicle level hazards and SOTIF requirements. In reality, there may be further hazards and derived requirements, which may involve also other aspects of the lateral motion, a such as acceleration and jerk, as well as the longitudinal motion of the vehicle, e.g., for overtaking maneuvers.

\section{SOTIF Background}\label{Sec:SOTIFBackground}

\subsection{Aspects of Safety}\label{Sec:SafetyAspects}

The \emph{safety} of a technical system is commonly defined as the absence of unreasonable risk for the health and life of humans \cite{ISO26262:2018}. A reasonable level of risk has to account for the fact that most technological systems, including ADSs, can never be entirely safe. Safety risks of E/E systems can be categorized according to the source of potential hazards into
\begin{itemize}
    \item[1)] hazards that occur due to \emph{malfunctioning behavior} (Functional Safety), addressed by the ISO\,26262 \cite{ISO26262:2018} standard, and
    \item[2)] hazards that occur due to \emph{functional insufficiencies} (Safety of the Intended Functionality, SOTIF), addressed by the ISO\,21448 \cite{ISO21448:2022} standard.
\end{itemize}
Functional safety and SOTIF are distinct and complementary aspects of safety that both need to be accounted for in the development of an ADS \cite{Con18}. The exclusive focus of this paper, however, is SOTIF.

\subsection{SOTIF}\label{Sec:SOTIF}

\begin{figure}[h]
\begin{center}
\includegraphics[width=1.0\textwidth]{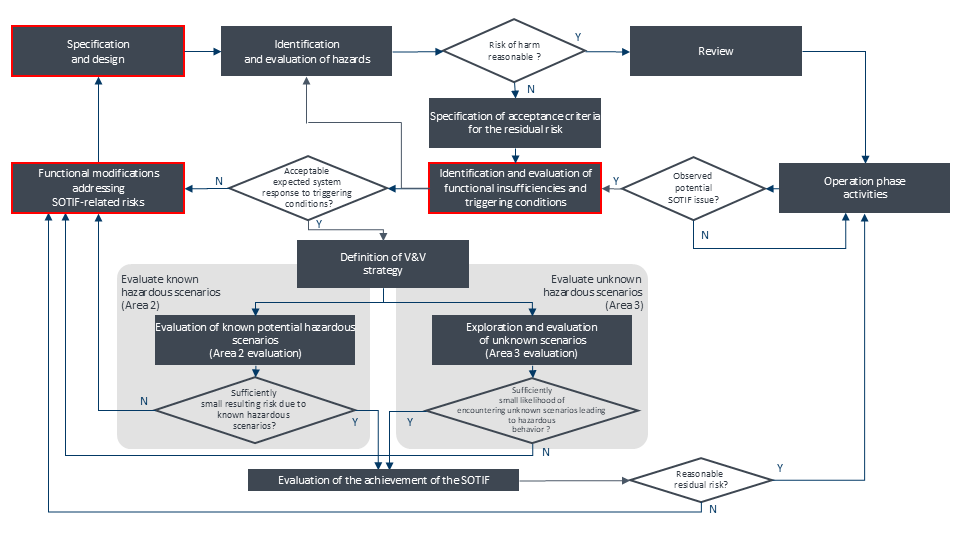}
\end{center}
\vspace*{-0.7cm}
\caption{Overview of the SOTIF lifecycle \cite{ISO21448:2022} (Red frames indicate the activities covered in this paper).\label{Fig:SOTIFLifecycle}}
\end{figure}

The international standard ISO\,21448 'Road vehicles - Safety of the intended functionality' published in 2022 defines SOTIF as the absence of unreasonable risk due to hazards resulting from \emph{functional insufficiencies} of the intended functionality, or its implementation \cite{ISO21448:2022}. As a basic framework, it defines an iterative lifecycle with several activities that are deemed adequate for developing an ADS that satisfies the SOTIF, as shown in Figure \ref{Fig:SOTIFLifecycle}.

\subsection{SOTIF-related Element out of Context (SOTIF-EooC)} \label{Sec:SOTIF-EooC}

As a consequence of the collaborative development typical for the automotive industry, a supplier company often provides a component that will be integrated by the OEM into a SOTIF-related system. Under this scenario, at the time of developing the component, the the supplier may only know certain aspects of the overarching system. One way of facilitating the achievement of SOTIF in such a context is to treat the component as a \emph{SOTIF-related Element out of Context (SOTIF-EooC)}, according to ISO\,21448 \cite[Clause 4.4.3]{ISO21448:2022}.   

For a component treated as a SOTIF-EooC, assumptions can be made regarding its use within the whole system and its contribution to the intended functionality. Based on this, assumptions about the SOTIF-related \emph{functional insufficiencies} (FIs) and their allowed target rate of occurrence can be derived. These assumptions need to be documented. They are used as inputs for the subsequent development of the SOTIF-EooC by the supplier. For the SOTIF-EooC, the identified \emph{triggering conditions} (TCs) of the component and their resulting FIs are documented together with the assumptions of use for the component (a.k.a. integration requirements).
When the OEM integrates the SOTIF-EooC, the validity of the assumptions or integration requirments is established by \mbox{SOTIF} activities in the context of whole vehicle-level functionalities, according to ISO\,21448, Clause 4.4.3 \cite{ISO21448:2022}.

In this paper, the MPC-TP is treated as a SOTIF-EooC. 
In fact, the MPC-TP may be used as part of different driving automation systems, namely E$^2$LCACC, ALCA, and HP. Assumptions regarding the overall systems, the integration of the MPC-TP and the contributions of the MPC-TP to the intended functionality have been discussed in Sections \ref{Sec:Specification} and \ref{Sec:DevelopmentScope}.
The identification of TCs of the MPC and their resulting FIs are discussed in Section \ref{Sec:SOTIFAnalysis}. Naturally, their analysis leads to an improved architecture and refined specifications, which are presented in Section \ref{Sec:RefinedSpecDesign}.

% The SOTIF approach also covers the dangers of reasonably foreseeable misuse by the driver or operator of the vehicle. However, foreseeable misuse is not considered further in this paper, as it is not central to the MPC-TP as a SOTIF-EooC. 

\subsection{Functional Insufficiencies}\label{Sec:FIs}

\begin{wrapfigure}{r}{0.55\textwidth}
	\vspace*{-0.3cm}
	\includegraphics[width=0.5\textwidth]{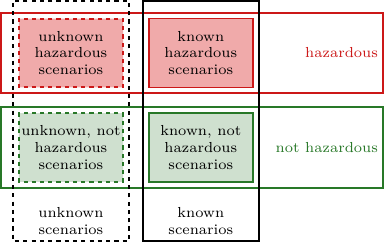}
	\vspace*{-0.2cm}
\caption{SOTIF scenario categories.\label{Fig:ScenarioCategories}}
	\vspace*{-0.7cm}
\end{wrapfigure}

FIs lead to hazards only in specific (driving) scenarios, called the \emph{hazardous scenarios}, as opposed to \emph{non-hazardous scenarios}. Conceptually, these scenarios are further sub-divided into \emph{known} and \emph{unknown scenarios}. This categorization of scenarios is illustrated in Figure \ref{Fig:ScenarioCategories}. Hazardous scenarios may be known or unknown at the start of the \mbox{SOTIF} analysis. The overall goal of SOTIF is to improve the specification and the design of the system, or the considered element, with respect to reducing the number of (known and unknown) hazardous scenarios. 

%\begin{figure}[h]
%\begin{center}
%\includegraphics[width=0.5\textwidth]{figures/HazardousScenarios.pdf}
%\end{center}
%\vspace*{-0.7cm}
%\caption{SOTIF scenario categories.\label{Fig:ScenarioCategories}}
%\end{figure}

FIs may exist at the system level, e.g., the E\textsuperscript{2}LCACC system, or at the element level, e.g., the 'Sense', 'Plan', or 'Act' parts of the system. Since FIs can lead to hazardous behaviour, their identification and the evaluation of their effects are key activities in the development of a SOTIF related item \cite[Clause 7]{ISO21448:2022}. ISO\,21448 subdivides FIs into
\begin{itemize}
    \item \emph{specification insufficiencies}, which refer to a (possibly incomplete) specification contributing to hazardous behavior, and
    \item \emph{performance insufficiencies}, which refer to limitations of the technical capabilities contributing to hazardous behavior \cite{ISO21448:2022}.
\end{itemize} 
% Specification insufficiency := Specification, possibly incomplete, contributing to either a hazardous behavior or an inability to prevent or detect and mitigate a reasonably foreseeable indirect misuse when activated by one or more triggering conditions \cite{ISO21448:2022}. 
% Performance insufficiency := Limitation of the technical capability contributing to a hazardous behavior or inability to prevent or detect and mitigate reasonably foreseeable indirect misuse when activated by one or more triggering conditions \cite{ISO21448:2022}.

Examples of specification insufficiencies include an incomplete specification of the E\textsuperscript{2}LCACC's headway distance resulting in the ego vehicle not keeping a safe distance to the vehicle in front or the inability of the HP to handle uncommon freeway road signs that were not part of its specification and thus cannot be processed appropriately by the system.
Examples of performance insufficiencies include limitations of technical capabilities of the ALCA such as limited calculation performance, limited perception range of its front radar sensor, or limited actuation capabilities of the foundational steering system  \cite{ISO21448:2022}.

As this paper is concerned with an MPC-TP for different driving functions, the focus is on FIs at element level, in this case, of a ``Plan'' component. 

% Depending on the system architecture, ISO\,21448 \cite{ISO21448:2022} further classifies potential functional insufficiencies into \emph{single-point functional insufficiencies} and \emph{multiple-point functional insufficiencies}.

\subsection{Triggering Conditions}\label{Sec:ScenarionsAndTrigConds}

Hazardous scenarios can be decomposed into the following elements: (1) A \emph{traffic scenario} with a \emph{triggering condition}, (2) which leads to a \emph{hazardous behavior} of the ego vehicle, (3) which may cause a \emph{hazard} in combination with the surrounding traffic or environment. Hence a TC is defined as a specific condition of a scenario that initiates a subsequent system reaction contributing to hazardous behavior \cite{ISO21448:2022}(simplified). 
%\textcolor{red}{Direct quote? Then we need to add quotation marks.--> quotation adapted such that it does not appear like a direct quote)}

\begin{figure}[h]
\begin{center}
\includegraphics[width=1.0\textwidth]{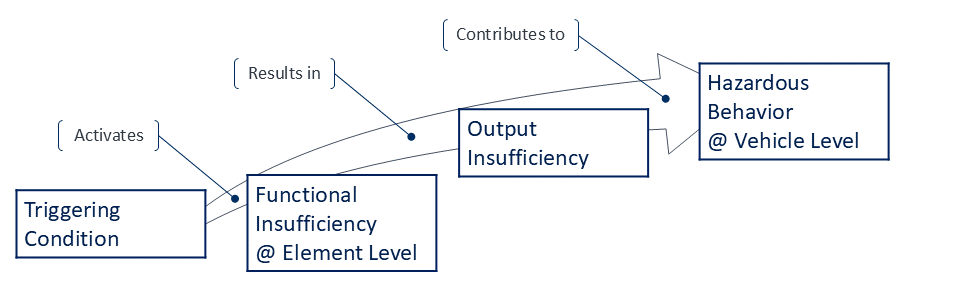}
\end{center}
\vspace*{-0.7cm}
\caption{Causal chain for hazardous behavior on vehicle level.\label{Fig:CausalChain}}
\end{figure}

TCs and FIs are interrelated: A TC of a scenario activates a FI, which results in a subsequent system reaction. Figure \ref{Fig:CausalChain} illustrates the activation of FI at the element level by a TC and the resulting hazardous behavior. In general, the SOTIF standard also addresses reasonably foreseeable misuse. This aspect however, remains outside the scope of this paper.

\begin{figure}[h]
\begin{center}
\includegraphics[width=1.0\textwidth]{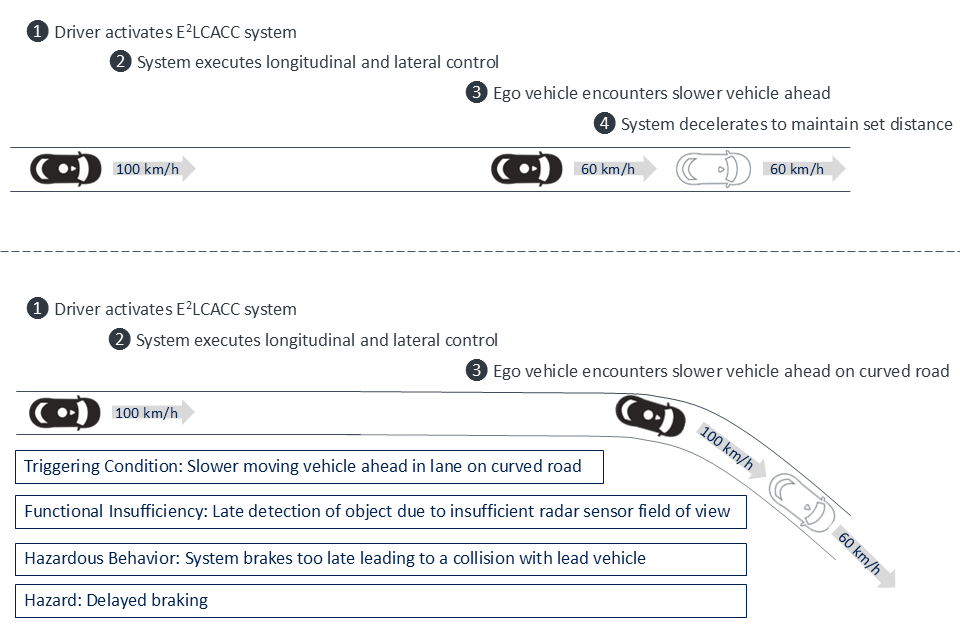}
\end{center}
\vspace*{-0.7cm}
\caption{Scenarios with (bottom) and without (top) TC.  \label{Fig:TCFIExample}}
\end{figure}

For an example, assume an activated E\textsuperscript{2}LCACC system that uses a radar sensor with an insufficient field of view in cased of curved roads. The upper part of Figure \ref{Fig:TCFIExample} shows a scenario without a TC. A FI is not being activated. The E\textsuperscript{2}LCACC system works as intended, i.e., the scenario is considered not hazardous.  The lower part of Figure \ref{Fig:TCFIExample} shows a second, hazardous scenario. Here, the TC is a slower moving vehicle ahead in lane on a curved road. In this case, the assumed FI at the sensor is activated, resulting in a late detection of the leading vehicle. As a result, the system brakes too late, resulting in a rear-end collision with the leading vehicle.   

% \textcolor{red}{[Idea: Use a planning example instead]}

\vspace*{-0.3cm}
\section{SOTIF Analysis for the MPC-TP}\label{Sec:SOTIFAnalysis}

\subsection{Identification and Evaluation of Functional Insufficiencies and Triggering Conditions}\label{Sec:FunInsufficiencies}

%It may be based on the 'Specification and Design', on the results of the 'Identification and evaluation of hazards' (a.k.a. SOTIF-HARA), as well as on already known potential functional insufficiencies or potential triggering conditions. 
ISO\,21448 suggests a list of specific methods to analyze potential FIs and TCs \cite[Table 4]{ISO21448:2022}. Table \ref{Tab:TCFIMethods1} shows the list of methods that are selected as appropriate to derive FIs and/or TCs of the MPC-TP, and the ones that are considered further in the remainder of this paper.

\begin{table}[h]
\renewcommand{\arraystretch}{1.5}
\begin{tabular}{|p{0.2cm}p{9.3cm}|p{1.3cm}|} 
  \hline
  \multicolumn{2}{|l|}{\textbf{Method}} &  \textbf{Section}\\\hline\hline
 \textcolor{gray}{A)} &  \textcolor{gray}{Analysis of requirements} &\\
	B) & Analysis of the ODD, use cases, and scenarios & \multicolumn{1}{c|}{\ref{Sec:AOUCS}}\\
%  C) & Analysis of accident statistics\\ 
	D) & Analysis of boundary values & \multicolumn{1}{c|}{\ref{Sec:ABV}}\\ 
 %  E) &  Analysis of equivalence classes \\
	F) & Analysis of functional dependencies & \multicolumn{1}{c|}{\ref{Sec:AFD}}\\
	G) & Analysis of common triggering conditions & \multicolumn{1}{c|}{\ref{Sec:ACTC}}\\ 
 \textcolor{gray}{H)} & \textcolor{gray}{Analysis of potential triggering conditions from field experience and lessons learnt} &\\ 
	I) & Analysis of system architecture (including redundancies) & \multicolumn{1}{c|}{\ref{Sec:ASA}}\\
%  J) & Analysis of design of the sensors and potential technology limitations\\ 
	K) & Analysis of algorithms and their output or decisions & \multicolumn{1}{c|}{\ref{Sec:AAOD}}\\ 
%  L) & Analysis of system ageing\\ 
%  M) & Analysis of possible environmental changes over vehicle operational lifetime\\
    \textcolor{gray}{N)} & \textcolor{gray}{Analysis of external and internal interfaces} & \\ 
    \textcolor{gray}{O)} & \textcolor{gray}{Analysis of design of the actuators and potential limitations} &\\ 
%  P) & Analysis of accident scenarios\\
%  Q) & Analysis of reasonably foreseeable misuse\\ 
 \hline
\end{tabular}
\caption{List of recommended methods in ISO\,21448 to analyze FIs and TCs that are deemed as relevant for the MPC-TP. Methods in gray are not considered further in this case study.\label{Tab:TCFIMethods1}}
\end{table}

For the analysis of the ``Plan'' component of a system, ISO\,21448 \cite[Section 7.3.2]{ISO21448:2022} provides an additional, specific list of methods that can be used for finding FIs and TCs. This list is provided in Table \ref{Tab:TCFIMethods2}. The categories considered for the SOTIF analysis in this paper are shown in black, the ones not considered further are shown in gray.

\begin{table}[h]
\renewcommand{\arraystretch}{1.5}
\begin{tabular}{|p{9.5cm}|p{1.3cm}|} 
  \hline
  \textbf{Method} &  \textbf{Section}\\\hline\hline
	Environment and location & \multicolumn{1}{c|}{\ref{Sec:EAL}}\\ 
	\textcolor{gray}{Road infrastructure} & \\ 
	\textcolor{gray}{Urban or rural infrastructure} & \\ 
	\textcolor{gray}{Highway infrastructure} & \\ 
	\textcolor{gray}{Driver or user behavior} & \\ % (incl.\ reasonably foreseeable misuse) \\ 
	\textcolor{gray}{Potential behavior of other drivers or road users} & \\ 
	\textcolor{gray}{Driving scenario} & \\ % (e.g., a construction site, an accident, a traffic jam with emergency corridor, vehicle driving in the wrong direction) \\
	\textcolor{gray}{Known planning algorithm limitations} & \\ % (e.g., inability to handle possible scenarios, or non-deterministic behavior) \\ 
	\textcolor{gray}{Known insufficiencies of the specification of machine learning} & \\ 
	%Known insufficiencies of the measurement data for machine learning \\
    \textcolor{gray}{Known functional insufficiencies and functional improvements} & \\ 
 \hline
\end{tabular}
\caption{List of recommended methods in ISO\,21448 to analyze FIs and TCs specifically for planning modules. Methods in gray are not considered further in this case study.\label{Tab:TCFIMethods2}}
\end{table}

This section describes the application of the selected analysis methods, and the FIs and potential TCs that have been identified. Multiple methods may lead to the same results, in which case the corresponding FIs and TCs are identified with the same number.

\subsection{Analysis of ODD, Use Cases, and Scenarios}\label{Sec:AOUCS}

This analysis method involves a close examination of the ODD specifications in (\Specref{Spec:ODD}), as well as predictable use cases and scenarios that may lead to potential TCs and FIs of the MPC-TP. Using this approach, the following items have been identified for our case study:

In congestion situations (\TCref{TC:AOUCS01}), the MPC-TP does not consider the formation of a rescue alley for police or ambulances (\FIref{FI:AOUCS01}). Conversely, the formation of a rescue alley by other vehicles or the presence  of police or ambulances in congestion situations may lead the MPC-TP to perform unusual, illegal, counter-intuitive or even dangerous maneuvers (\FIref{FI:DangerousManeuver}).
%Hence, a congestion situation must be detected by the Map Fusion module. Then the MPC-TP switches to a congestion mode, that is able to manage the formation of rescue alleys.

In obstacle avoidance scenarios (\TCref{TC:AOUCS02}), the MPC-TP does not allow emergency maneuvers without a proper lane change. For example, it is not possible to avoid an obstacle by driving in between two lanes or by using the shoulder (\FIref{FI:AOUCS02}). Conversely, the emergency obstacle avoidance by other vehicles may lead the MPC-TP to perform counter-intuitive or even dangerous maneuvers (\FIref{FI:DangerousManeuver}).
% In emergency cases, the MPC-TP shall be able to switch to an obstacle avoidance mode, in which the lane keeping constraints are relaxed, but the collision constraints remain intact.

There is no speed limit on certain sections of the Autobahn. Suppose there is blocking obstacle on the road, such as the rear end of a congestion. If the obstacle is at standstill shortly beyond the prediction horizon of the MPC-TP (\TCref{TC:AOUCS03}) and an obstacle avoidance maneuver is not possible, e.g., by performing a lane change. If the EgoV exceeds a certain speed $v_{\max}$, this can lead to a rear-end collision because the prediction horizon $N_{\mathrm{p}}$ is not long enough for the EgoV to come to a full stop within it (\FIref{FI:AOUCS03}). 
%Since $N_{\mathrm{p}}$ is static, it must be chosen in accordance with the maximum deceleration $a_{\mathrm{lon},\min}$ and the maximum velocity $v_{\max}$ in (S01j):
%\begin{equation}\label{Equ:MaxVel} v_{\max}=N_{\mathrm{p}}\,a_{\mathrm{lon},\min}\,t_{\mathrm{s}}\;.
%\end{equation}
%The maximum deceleration $a_{\mathrm{lon},\min}\leq 0$ is the worst-case (highest) deceleration, depending on current road conditions. It depends on the worst-case values of the ODD, in particular the friction value $\mu$. In addition to existing constraints, the MPC-TP must consider the speed limit in \eqref{Equ:MaxVel} above.

The design of the current MPC-TP is based on clearly defined traffic rules. It assumes, that surrounding vehicles either have the right-of-way, and are thus considered as obstacles, or they have to yield, and can thus be ignored. However, there are situations where the right-of-way is unclear, such as at the end of two merging lanes (\TCref{TC:AOUCS04}a). Without further measures, this may lead to an overly conservative driving behavior (\FIref{FI:AOUCS04}a), and even a complete deadlock for the EgoV (\FIref{FI:AOUCS04}b).

On freeways, cars typically only drive forwards. However, in exceptional situations, e.g., during congestion or due to police or ambulance operations, it may be necessary to perform reverse maneuvers. Moreover, surrounding vehicles may drive backwards (\TCref{TC:AOUCS05}), in which case the MPC-TP may become infeasible (\FIref{FI:Infeasible}) if the EgoV is at rest or slowly moving forwards.

In the case of a wrong way driver (\TCref{TC:AOUCS06}), the MPC-TP may be unable to compute a feasible trajectory (\FIref{FI:Infeasible}). Or it may lead the MPC-TP to perform counter-intuitive or even dangerous maneuvers (\FIref{FI:DangerousManeuver}).

The MPC-TP shall be able handle situations with an unexpected, sudden change of road conditions. Important examples include a dropping friction coefficient, e.g., due to water / hydroplaning (\TCref{TC:AOUCS07}a), snow / ice (\TCref{TC:AOUCS07}b), or oil (\TCref{TC:AOUCS07}c), or an irregular surface, e.g., due to gravel or stones (\TCref{TC:AOUCS08}a), pot holes (\TCref{TC:AOUCS08}b), or road bumps (\TCref{TC:AOUCS08}c). In such situations, the trajectory generated by the MPC-TP may actually be not drivable, in the sense of (\SRref{SR:Drivable}), e.g., because of the limitations of the LDBM using a linear tire model (\FIref{FI:Drivable}).
%Analysis of ODD: Friction coefficient may be lower than expected temperatures below 4 degrees are removed from the ODD and there is a reliable friction coefficient estimator (parameter estimator) that give the friction coefficient $\mu$ estimation within a given tolerance $\Delta\mu$, MPC model switches to dynamic bicycle model instead of kinematic bicycle and uses $\mu-\Delta\mu$.

\subsection{Analysis of System Architecture (Including Redundancies)}\label{Sec:ASA}

This analysis is based on the system architecture, as depicted in Figure \ref{Fig:Architecture}. The goal is to identify potential FIs and TCs based on the overall behavior of the entire system.

The receding-horizon implementation causes the MPC-TP to act also as a feedback controller for the vehicle, which is a well-known feature of MPC \cite{Mayne:2000,Werling:2011}. The interoperation of the MPC-TP with the tracking controller (TC) may thus lead to undesired bahavior, such as oscillations (\FIref{FI:ASA01}), or even instability of the EgoV motion. This may be triggered, for example, by external disturbances such as a side wind gust, e.g., on a bridge deck (\TCref{TC:ASA01}), or irregular, non-continuous lane markings, e.g., due to suddenly disappearing (\TCref{TC:ASA02}a), dirty (\TCref{TC:ASA02}b), broken (\TCref{TC:ASA02}c), or covered (\TCref{TC:ASA02}d) lane markings.

Difficulties may also appear in situations where the right-of-way is unclear (\TCref{TC:AOUCS04}), or when other traffic participants do not strictly adhere to the traffic rules (\TCref{TC:ASA05}). Their interaction with the MPC-TP may then lead to an overly conservative behavior of the EgoV (\FIref{FI:AOUCS04}), possibly resulting in a poor progress on the road (\FIref{FI:AOUCS04}a), or even a deadlock (\FIref{FI:AOUCS04}b).

The cut-in of a faster TgtV in front of the EgoV (\TCref{TC:ASA04}) may lead to a violation of the front safety distance, and hence the inability of the MPC-TP to generate a feasible solution (\FIref{FI:Infeasible}). This can be resolved by relaxing the minimum safety distance constraint in case of a leading vehicle with a higher velocity. Even if the safety distance is not violated, the cut-in of a faster TgtV (\TCref{TC:ASA04}) may still lead to an overly conservative braking maneuver on behalf of the EgoV (\FIref{FI:AOUCS04}c).

The MPC-TP does not consider a trailing vehicle driving closely behind the EgoV in the same lane (\TCref{TC:ASA06}). According to (\Specref{Spec:MPC}d), even though it is the responsibility of the trailing vehicle to keep a sufficient safety distance, sudden braking maneuvers may lead to a collision with the trailing vehicle and are thus unsafe (\FIref{FI:ASA03}).

\subsection{Analysis of Functional Dependencies}\label{Sec:AFD}

In the context of this case study, functional dependencies are interpreted as other modules that are actually or potentially connected with the MPC-TP, be it upstream or downstream. The MPC-TP needs to cope with challenges arising from the limitations of these modules. With this interpretation of functional dependencies, TCs and FIs identified with this analysis method may overlap with those identified by analyzing the system architecture (in Section \ref{Sec:ASA}), in the context of the entire system.

Dynamic obstacles suddenly appearing in the field of view of the system (\TCref{TC:AFD01}), may overwhelm the ability of the MPC-TP to solve the OCP in order to compute a new trajectory in due time (\FIref{FI:InsufficientTime}), violating (\SRref{SR:Timing}), or to find a feasible trajectory at all (\FIref{FI:Infeasible}), violating (\SRref{SR:CollisionFree}). Examples of this group of FIs are a previously shadowed traffic participant (\TCref{TC:AFD01}a), an object dropping from a bridge or a truck (\TCref{TC:AFD01}b), or a large animal (e.g., a wild boar) crossing the freeway (\TCref{TC:AFD01}c).

At points of sudden traffic rule changes (\TCref{TC:AFD02}), such as the beginning of a speed limit area or a no passing zone, the MPC-TP may also unable to compute a feasible trajectory (\FIref{FI:Infeasible}), or may be unable to solve the OCP within the allotted time (\FIref{FI:InsufficientTime}).

\subsection{Analysis of Algorithms and Their Outputs or Decisions}\label{Sec:AAOD}

In the context of the MPC-TP, this analysis pertains to the formulation of the OCP \eqref{Equ:MSSP} and the numerical optimization algorithm, the decisions taken by the MPC-TP, as well as the reference trajectory and the exit flag as its outputs.

The OCP is based on the LDBM, whose linear tire model may represent a poor approximation of a real vehicle if essential model assumptions are violated (\FIref{FI:AAOD01}). The validity of the linear tire model is already supported by corresponding constraints $\mathbb{X}^{(\mathrm{s})}_{t}$ (see Section  \ref{Sec:MPC-TP}). The other main assumption of the LDBM is a non-negligible roll motion of the vehicle. It can be violated by an extremely asymmetric load (\TCref{TC:AAOD01}a), a highly dynamic cornering maneuver (\TCref{TC:AAOD01}b), or a high position of the CoG (\TCref{TC:AAOD01}c).

Potential problems arising from the interoperation of the MPC-TP with the TC and from the lack of interaction-awareness have already been discussed (see Section \ref{Sec:ASA}). In addition, the MPC-TP may cause a lack of string stability in a chain of vehicles (\TCref{TC:StringStability}). If the leading brakes  abruptly, this can lead to the build-up of an increasing wave of braking maneuvers by all following vehicles. This string instability may be fostered by the MPC-TP in the middle of a chain of human-driven or automated vehicle, which may lead to collisions (\FIref{FI:StringStability}), if the MPC-TP is not properly tuned.

The outputs of the MPC-TP include the reference trajectory and the solver exit flag (see Figure \ref{Fig:Architecture}).

Without software or hardware faults, the computed reference trajectory will be consistent, admissible, drivable, and collision-free, if the set $\mathcal{T}$ of such trajectories is non-empty. However, due to the finite prediction time of $T_{\mathrm{p}}=6\,\mathrm{s}$, the MPC-TP may lead the vehicle to an (unsafe) situation that cannot be resolved by a future trajectory, i.e., where $\mathcal{T}$ becomes empty. In MPC language, this is known as the lack of \emph{recursive feasibility} (\FIref{FI:RecursiveFeasibility}). A possible TC is the fast approach of a slowly driving TgtV in the front of the EgoV, where the TgtV cannot be avoided by a lateral maneuver (\TCref{TC:SlowLeadingTgtV}).

The solver exit flag itself might provide an indication of a triggered FI. In particular, all exit flag values $q_{\mathrm{exit}}\neq 1$ should be considered carefully:
\begin{itemize}
    \item $q_{\mathrm{exit}}=2$ indicates that the MPC may exceed the allotted computation time (\FIref{FI:InsufficientTime}), but produces a feasible trajectory. This could be triggered by rapidly changing traffic situations (\TCref{TC:AAOD02}) or by suddenly appearing obstacles (\TCref{TC:AFD01}). 
    \item $q_{\mathrm{exit}}=3$ means that the MPC-TP is unable to generate a feasible (collision-free) trajectory for the given constraints, because $\mathcal{T}$ is empty (\FIref{FI:Infeasible}). For example, this can be triggered by (\TCref{TC:AOUCS05}), (\TCref{TC:AOUCS06}), (\TCref{TC:AOUCS07}), (\TCref{TC:AOUCS08}), (\TCref{TC:ASA04}), (\TCref{TC:AFD01}),(\TCref{TC:AFD02}).
    \item $q_{\mathrm{exit}}=4$ means that the solver is unable to find a solution (\FIref{FI:NoSolution}), even though $\mathcal{T}$ may be non-empty. As software and hardware faults are out of scope, this must be due to unpredictable numerical problems (e.g., ill-conditioning, floating point arithmetics). To the best of the authors' knowledge, there are no known specific triggering conditions for this case.
\end{itemize}

%\subsection{Analysis of External and Internal Interfaces}

\subsection{Analysis of Boundary Values}\label{Sec:ABV}

For the analysis of boundary values, the MPC-TP is examined in the context of scenarios where relevant parameters take on extreme values. Parameters considered in the analysis include
\begin{itemize}
    \item \emph{ODD and road parameters}, i.e., the road bank angle, the friction coefficient $\mu$, the road curvature, the change of curvature, and the wind speed;
    \item \emph{maneuver parameters}, i.e., the reference velocity, maximum lateral motion (e.g., due to a sudden lane change or obstacle avoidance), maximum longitudinal motion (e.g., due to emergency braking or maximum acceleration);
    \item \emph{vehicle parameters}, i.e., the vehicle mass ($m$), the moment of inertia ($I_{z}$), the position of the CoG ($x_{\mathrm{cog}}$, $y_{\mathrm{cog}}$, $z_{\mathrm{cog}}$), the cornering stiffness of the front and the the rear tires ($C_{\alpha,\mathrm{f}}$, $C_{\alpha,\mathrm{r}}$), and all chassis and suspension parameters of the specified vehicle class.
\end{itemize}

The analysis is based on extensive simulations using a high-fidelity vehicle model, e.g, in CarMaker. The simulations scenarios aim to cover possible combinations of extreme values of the above parameters. The results are checked against the requirements (\SRref{SR:Timing}) - (\SRref{SR:CollisionFree}).

To this end, a database of artificial test scenarios covering \emph{corner cases}, is generated, based on different incarnations of the \emph{maneuver parameters}. Specifically for the reference velocity, scenarios are separated into \emph{low-speed} ($v\leq v_{\mathrm{low}}=5\,\frac{\mathrm{km}}{\mathrm{h}}$), \emph{medium-speed} ($v\leq v_{\mathrm{med}}=50\,\frac{\mathrm{km}}{\mathrm{h}}$), and \emph{high-speed} ($v\leq v_{\mathrm{high}}=130\,\frac{\mathrm{km}}{\mathrm{h}}$) categories. For road curvature and change of curvature, only scenarios that (at least partially) involve maximum values are considered. The generation process of corner cases can also be automated, as proposed in Lubiniecki et al.\ \cite{LubinieckiEtAl:2020}.

The database of corner cases is then simulated for extreme values of the \emph{ODD and road parameters} and the \emph{vehicle parameters}. Due to the combinatorial nature, the number of simulation scenarios thus becomes intractable. To regain tractability, only \emph{pairs} of extreme parameter values are included in the boundary value analysis, while all other parameters are set to their default or regular values. For some parameters, such as the height of the CoG $z_{\mathrm{cog}}$, instead of a default or regular values, it is more reasonable to consider only the maximum value in all simulations. Critical cases are expected to result from certain parameter combinations (in some or all corner cases), such as
\begin{itemize}
    \item the maximum side wind speed $v_{\mathrm{wind},\max} = 8\,\mathrm{bft}$ and the maximum bank angle $\beta_{\max} = 8^{\circ}$ (\TCref{TC:CommonTrigger1}),
    \item the minimum friction value $\mu_{\min}=0.15$ and the minimum tire cornering stiffness $C_{\alpha,\mathrm{f},\min}=60\,\frac{\mathrm{kN}}{\mathrm{rad}}$ (\TCref{TC:CommonTrigger2}),
    \item maximum front tire cornering stiffness $C_{\alpha,\mathrm{f},\max}=200\,\frac{\mathrm{kN}}{\mathrm{rad}}$ and the minimum rear tire cornering stiffness $C_{\alpha,\mathrm{r},\min}=60\,\frac{\mathrm{kN}}{\mathrm{rad}}$, and vice versa (\TCref{TC:CommonTrigger3}).
\end{itemize}

These TCs may lead the MPC-TP to provide trajectories that are not drivable by a specific vehicle model or configuration (\FIref{FI:Drivable}). However, these are potential TCs and FIs, and their existence and relevance has to be confirmed by actually performing a large-scale, high-fidelity simulation study.

\subsection{Analysis of Common Triggering Conditions}\label{Sec:ACTC}

Generally, common TCs are collected from prior experiences with the development of similar systems. They may be available, for example, by internal reports, industry standards, or other publicly documents. The TCs identified in this way may, of course, overlap with TCs identified using other methods.

The authors are currently not aware of a database or any other source for common TCs for a trajectory planner. However, applicable TCs for other ADSs from the relevant literature have been taken into account in this case study. In particular, selected TCs for a LKA system from \cite{9872242} can analyzed with respect to their ability to challenge the MPC-TP and to trigger potential FIs:
\begin{itemize}
    \item lane keeping / lane changing scenario with suddenly disappearing lane markings (\TCref{TC:ASA02}a);
    \item lane keeping / lane changing scenario with dirty, broken, or covered lane markings (\TCref{TC:ASA02}b,c,d);
    \item lane keeping / lane changing scenario with unrecognizable lane markings due to direct sunlight or change of lighting conditions, e.g., when entering a tunnel (\TCref{TC:ACTC03}a,b);
    \item insufficient visibility of lane markings due to low visibility, e.g., due to snow / ice or fog (\TCref{TC:ACTC04}a,b);
    \item  infrastructure items that could be misinterpreted as lane markings, e.g., curbs or dividers (\TCref{TC:ACTC05}a,b).
\end{itemize}
Other common TCs for a trajectory planner that can been identified from the existing literature include
\begin{itemize}
    \item side wind gusts, e.g., on a bridge deck (\TCref{TC:ASA01}a),
    \item an uneven road surface, due to potholes (\TCref{TC:AOUCS08}b) or road bumps (\TCref{TC:AOUCS08}c),
    \item a slippery road surface, e.g., due to a wet road with foilage (\TCref{TC:ACTC09}a) or a snowy / icy road (\TCref{TC:ACTC09}b).
\end{itemize}

For many of the common TCs from literature it remains unclear however, whether they activate FIs in the MCP-TP or elsewhere in the system. This needs to be verified in simulations or real-world experiments.

\subsection{Environment and Location}\label{Sec:EAL}

The systematic analysis of the environment and location aims to identify relevant freeway scenarios and warning signs that may compromise the safety of the MPC-TP. By means of brainstorming, the following scenarios in the ODD can be identified as potential TCs in this context.

An increase or decrease in the number of lanes on the freeway may present a particularly challenging situation for the MPC-TP, including:
\begin{itemize}
    \item entering the freeway from an on-ramp (\TCref{TC:EAL01}),
    \item exiting the freeway via an off-ramp (\TCref{TC:EAL02}),
    \item lane ending / lane merging situations (\TCref{TC:AOUCS04}a),
    \item lane adding situations (\TCref{TC:EAL04}).
\end{itemize}

Similarly, situations with special vehicles or vehicles operating beyond traffic rules may pose particular difficulties for the MPC-TP, such as: 
\begin{itemize}
    \item a wrong way driver (\TCref{TC:AOUCS06}), 
    \item a reckless driver or person driving under-the-influence (\TCref{TC:EAL08}),
    \item an approaching emergency vehicle (\TCref{TC:EAL05}),
    \item a heavy transport or oversized vehicles (\TCref{TC:EAL06}),
    \item a vehicle parking on the freeway shoulder (\TCref{TC:EAL07}).
\end{itemize}
%Large animals suddenly appearing on the road (\TCref{TC:AFD01}c) or static or dynamic objects, such as dropped objects from a bridge or truck that block the road (\TCref{TC:AFD01}b) have been also identified in this brainstorming.

Furthermore, rare situations and uncommon road signs on the German Autobahn can be considered in the context of this analysis, leading to the following list of TCs:
\begin{itemize}
    \item uneven road surface (road sign 112), e.g., due to ruts or blow-ups (\TCref{TC:AOUCS08}d,e),
    \item an invisible / indetectable substance is substantially lowering the friction coefficient, e.g., due to an oil spillage (road sign 114) (\TCref{TC:AOUCS07}c),
    \item the freeway shoulder is not available (\TCref{TC:EAL09}) or the shoulder is to be used as an additional lane (road signs 223.1, 223.2, 223.3) (\TCref{TC:EAL10}),
    \item end of the freeway (road sign 330.1) (\TCref{TC:EAL11}) or a freeway closure (\TCref{TC:EAL12}),
    \item an unusually steep slope (road signs 108, 110) (\TCref{TC:EAL13}) or an usually high curvature (road signs 103-10, 103-20, 105-10, 105-20) of the road (\TCref{TC:EAL14}),
    \item dangerous side wind gusts (road sign 117) (\TCref{TC:ASA01}),
    %\item customized traffic signs (free text) in special situations
    \item potentially unknown traffic scenarios or traffic signs in other European countries (\TCref{TC:EAL15}).
\end{itemize}

All of the above TCs may affect the ability of the MPC-TP to solve the OCP and compute a trajectory in due time (\FIref{FI:InsufficientTime}), or to find a feasible (collision-free) trajectory at all (\FIref{FI:Infeasible}). Furthermore, they may cause the MPC-TP to provide trajectories which are not drivable and thus lead the EgoV to leave its assigned lane and/or collide with other vehicles or obstacles (\FIref{FI:Drivable}) or which may cause the vehicle to perform unusual, illegal, counter-intuitive or even dangerous maneuvers (\FIref{FI:DangerousManeuver}).

\subsection{Summary of Identified Triggering Conditions and Functional Insuffciencies}

Appendix \ref{Sec:Appendix} provides a table with an overview of all triggering conditions and functional insuffciencies that have been identified for the MPC-TP in this case study.

\section{Refined Concept for the Vehicle Level Control System and the MPC-TP}\label{Sec:RefinedSpecDesign}

As per the SOTIF lifecycle, the SOTIF-EooC analysis in Section \ref{Sec:SOTIFAnalysis} leads to a revision of the overall concept of the vehicle level control system and the MPC-TP, in particular. The results include a refinement of the MPC-TP specifications and design as well as the ODD, as explained in Sections \ref{Sec:RefinedSpecDesign} and \ref{Sec:RefinedODD}). As a further consequence, the integration requirements to the vehicle level control system are updated, as detailed in Section \ref{Sec:IntegrationReqs}. Some situations that could not be (fully) resolved as part of this case study are described in Sections \label{Sec:FurtherAnalysis} and \label{Sec:InteractiveMPC}.

\subsection{Refined Specification and Design of the MPC-TP}\label{Sec:RefinedSpecDesign}

\begin{figure}[h]
\begin{center}
\includegraphics[width=1.0\textwidth]{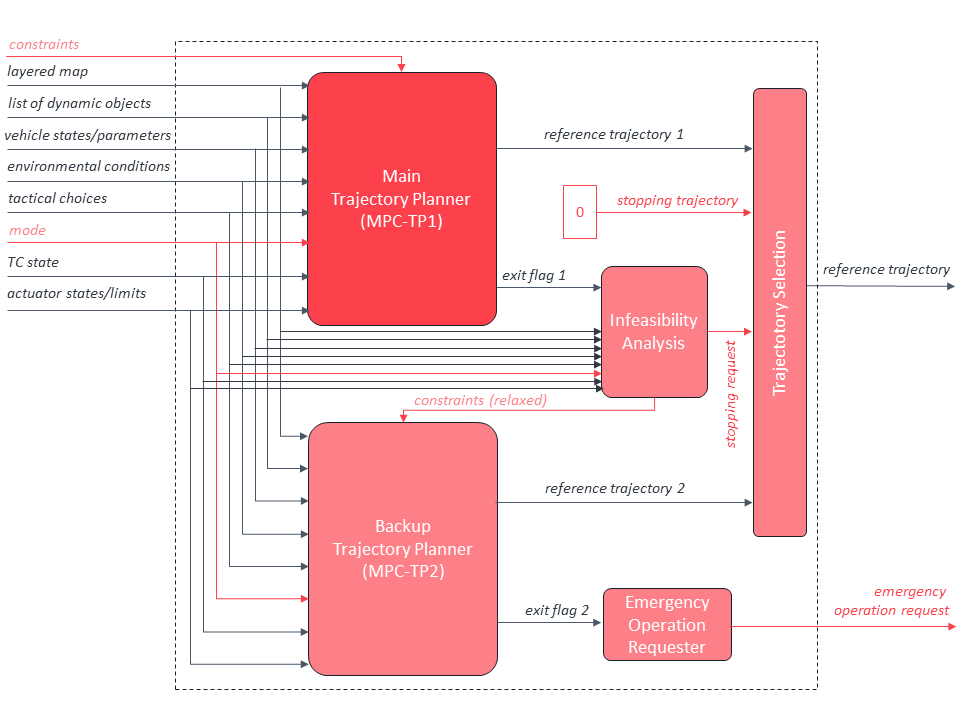}
\end{center}
\vspace*{-0.7cm}
\caption{Refined Architecture of the MPC-based trajectory planner. \label{Fig:RefinedArchitecture}}
\end{figure}

The refined system architecture based on the SOTIF analysis in Section \ref{Sec:SOTIFAnalysis} is depicted in Figure \ref{Fig:RefinedArchitecture}. The particular refinements of the specification and the design of the MPC-TP are discussed below.

\subsubsection{Design Improvements}\label{Sec:RefinedMPC}

%MC240424: IMHO not needed for this chapter 
%The following TCs have been considered based on known weak points of a trajectory planner in general:
%\begin{itemize}
%    \item corner cases in the vehicle parameters (vehicle mass, moment of inertia, position of the center of gravity, coefficient, cornering stiffness),
%    \item corner cases in the environment parameters (road friction, side wind, bank angle),
%    \item saturation of the actuator constraints (longitudinal and lateral acceleration, maximum steering angle and steering rate),
%    \item critical cases of the reference trajectory from the admissible set $\mathcal{T}$ (generally unknown).
%\end{itemize}
%Additionally, there are some known weak points of an MPC-based planner in particular:
%\begin{itemize}
%    \item exceeding the allotted computation time of $t_{\textrm{s}}=30\,\textrm{ms}$ (generally unknown, often due to rapidly changing reference trajectories),
%\end{itemize}

The SOTIF-EooC analysis has led to the following concepts to improve the architecture and algorithmic design of the MPC-TP.

An overly conservative behavior of the EgoV (\FIref{FI:AOUCS04}c) may result, e.g., due to the cut-in of a faster TgtV (\TCref{TC:ASA04}). This potential issue can be mitigated by a proper tuning of the MPC-TP.

A trailing TgtV not keeping a proper safety distance to the EgoV (\TCref{TC:ASA06}) may cause a collision in case of a sudden braking maneuvers of the EgoV (\FIref{FI:ASA03}). 

{\Spec\label{Spec:SuddenBraking}}a*) Even though this lies within the responsibility of the TgtV,  the EgoV shall perform sudden braking maneuvers only when strictly necessary and only to an appropriate extent. This can be accomplished by adding a strong penalty on the usage of high braking inputs in the OCP.

\Specref{Spec:SuddenBraking}b*) Furthermore, the distance to the trailing vehicle shall aslo be included in the tuning of the MPC-TP.

The interaction with other traffic participants if the right-of-way is unclear (\TCref{TC:AOUCS04}) or if they do not strictly adhere to the traffic rules (\TCref{TC:ASA05}) may lead to an overly conservative behavior of the EgoV (\FIref{FI:AOUCS04}). 

{\Spec\label{Spec:InteractionAwareMPC}}*) This should be properly resolved by using an interaction-aware MPC approach, which is the subject of ongoing research \cite{NairEtAl:2022,BencioliniEtAl:2023,ChenEtAl:2022,OliveiraEtAl:2023}. For the given architecture, the interaction-aware MPC also has to be designed in coordination with the Tactical Planner.

Finally, a lack of string stability may be fostered by the MPC-TP in a chain of human-driven or automated vehicles (\TCref{TC:StringStability}), which may potentially lead to collisions (\FIref{FI:StringStability}). The problem, however, is not MPC-specific and has been analyzed in the previous literature, including ACC systems \cite{SwaHed:1996,ShawHed:2007}. It shall be resolved by a proper tuning of the MPC-TP, which should be verified in extensive simulations and/or on-road tests.

\subsubsection{Adaptations to the Reference Trajectory}\label{Sec:ReactionHorizon}

\begin{figure}[h]
	\includegraphics[width=1.0\textwidth]{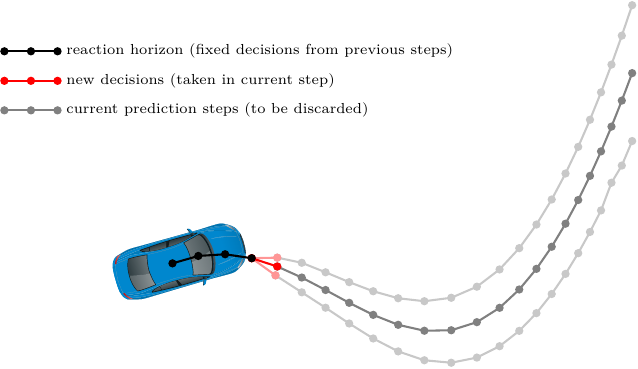}
\vspace*{-0.5cm}
\caption{Illustration of the reaction horizon.\label{Fig:DecisionHorizon}}
\end{figure}

The interoperation of the MPC-TP with the lower-level TC may lead to oscillations and even unstable behavior (\FIref{FI:ASA01}). 
\Specref{Spec:TrajectoryBody}g*) As a solution to this, the first $N_{\mathrm{r}}=4$ steps of the previously computed trajectory remain unchanged, i.e., only the subsequent inputs are newly computed. $N_{\mathrm{r}}$ is called the \emph{reaction horizon}. The concept is illustrated in Figure \ref{Fig:DecisionHorizon}. The initial condition for the OCP in (\ref{Equ:MSSP}c) thus becomes the predicted state $\hat{x}_{N_{\mathrm{r}}}$ from the previous OCP solution.

As a negative consequence of this modification, the reaction time of the MPC-TP is increased. This is relevant, for instance, if the EgoV has to react to sudden events in its environment, such as in an emergency braking or obstacle avoidance scenario. However, a reaction time of $T_{\mathrm{d}}=N_{\mathrm{r}}t_{\mathrm{s}}=200\,\mathrm{ms}$ is still superior to that of an average human driver, which is typically between $600\,\mathrm{ms}$ and $1,000\,\mathrm{ms}$, according to recent studies \cite{WolfeEtAl:2020}. Thus a reaction time of $200\,\mathrm{ms}$ is deemed to be sufficient for common road traffic.

\subsubsection{Potential Extension of Computation Time}\label{Sec:ComputationTime}

The regular MPC is designed for a maximum computation time of $t_{\mathrm{c}}=30\,\mathrm{ms}$. This is less than the given sampling time, $t_{\mathrm{s}}=50\,\mathrm{ms}$, in order to account for additional latencies, e.g., due to communication and actuation delays. Obviously, the choice of $t_{\mathrm{c}}$ should be based on generous computational margins and extensive testing. However, it is generally not possible to guarantee that the solver always provides a definite answer within this allotted computation time, i.e., that it finishes with an exit flag $q_{\mathrm{exit}} = 1,2,3$ in due time (\FIref{FI:InsufficientTime}). Several TCs have been identified as part of the SOTIF analysis in Section \ref{Sec:SOTIFAnalysis} that may represent a potential cause for an excess computation time, such as a suddenly appearing obstacle (\TCref{TC:AFD01}), sudden changes of traffic rules (\TCref{TC:AFD02}), or rapidly changing traffic situations (\TCref{TC:AAOD02}). A strategy to remedy this issue may look as follows:

Due to the concept of the reaction horizon, introduced in Section \ref{Sec:ReactionHorizon}, the computation time can be extended by multiple (in this case: $N_{\mathrm{r}}=4$) time steps. In occasional peaks, it can reach up to a theoretical maximum time of $\bar{t}_{\mathrm{c}}=t_{\mathrm{c}}+(N_{\mathrm{r}}-1)t_{\mathrm{s}}=180\,\mathrm{ms}$. After that, a new trajectory must definitely be provided. If multiple time steps are used to solve the OCP, the new decisions (shown in red in Figure \ref{Fig:DecisionHorizon}) comprise a corresponding number of time steps, which are now more than a single one.

Furthermore, the solver settings are adjusted such that the iterations can be halted after $t_{\mathrm{c}}$, if they have not terminated; or again after each $t_{\mathrm{c}}+k\,t_{\mathrm{s}}$, for $k=1,2,\dots,N_{\mathrm{r}}-1$. After each instance, the status of the current iterate can be checked. If the current exit flag equals to $q_{\mathrm{exit}} = 4$, the solver may continue its computations from the current iterate, until finally the ultimate maximum solver time $\bar{t}_{\mathrm{c}}$ is reached.

%If this happens, the safety logic switches to an Emergency Operation, for which further details are provided in Section \ref{Sec:BackupMPC}. If, however, the solver finishes with $q_{\mathrm{exit}} = 1,2,3$ at any of the steps before, then the MPC-TP either continues with the optimal trajectory (case $q_{\mathrm{exit}}=1$) or proceeds with the Infeasibility Analysis (cases $q_{\mathrm{exit}}=2,3$), described in Section \ref{Sec:InfeasibilityAnalysis}.

\subsubsection{Backup MPC}\label{Sec:BackupMPC}

The \emph{Backup MPC} (MPC-TP2) is a redundant implementation of the \emph{Primary MPC} (MPC-TP1), which has been discussed thus far. The Backup MPC has the same inputs and outputs as the Primary MPC, as shown in Figure \ref{Fig:RefinedArchitecture}). It is designed according to the same specifications and requirements, as described in Sections \ref{Sec:Specification} and \ref{Sec:DevelopmentScope}, except for the following differences:
\begin{itemize}
    \item The Backup MPC is utilized for safety and not passenger comfort. It has the ability to perform any maneuver that is physically possible. This means a different tuning of the weight parameters in the cost function and the traffic rule and comfort constraints $\mathbb{X}^{(\mathrm{t})}_{t}$ are removed from the OCP in (\ref{Equ:MSSP}f).
    \item Instead of a LDBM, which has a linear tire model, the Backup MPC shall include a NDBM with a Pacejka tire model. The NDBM provides a higher fidelity for the dynamics of the vehicle at the limits of handling, even though it may have a longer solution time.
    \item For additional redundancy, the Backup MPC shall apply a different numerical solver (different algorithm and/or software implementation) and it shall be running on an independent computational platform.
\end{itemize}

The Backup MPC is constantly running in a \emph{shadow mode}. During normal operation, when the Primary MPC is active, the solutions of the Backup MPC solutions are simply discarded. The trajectories of the Backup MPC are activated, or subsequently deactivated, according to a new module called \emph{Infeasibility Analysis}. Its logic is based on the exit flag of the Primary MPC $q_{\mathrm{exit}}^{\mathrm{(P)}}$, the exit flag of the Backup MPC $q_{\mathrm{exit}}^{\mathrm{(B)}}$, and various information from the perception modules; see Section \ref{Sec:InfeasibilityAnalysis} for further details. In particular, the Backup MPC shall be activated if the Primary MPC is unable to find a feasible solution (\FIref{FI:Infeasible}) under the following circumstances:
\begin{itemize}
    \item The friction coefficient suddenly drops by a value of $\mu_{\mathrm{d}}>0$ within a specified time interval $t_{\mathrm{d}}>0$, due to water, snow / ice, or oil (\TCref{TC:AOUCS07}), or an irregular surface, e.g., due to gravel or stones, pot holes, or road bumps (\TCref{TC:AOUCS08}).
    \item Static or dynamic obstacles suddenly appear close to the vehicle, e.g., due to a previously shadowed traffic participant (\TCref{TC:AFD01}a), a lost cargo or other object dropping from a bridge or a truck (\TCref{TC:AFD01}b), or a large animal (e.g., a wild boar) crossing the freeway (\TCref{TC:AFD01}c).
    \item There is a sudden change of traffic rules (\TCref{TC:AFD02}), such as the beginning of a speed limit zone or a no passing zone.
\end{itemize}

The Backup MPC remains active until the Primary MPC produces a optimal solution again (case $q_{\mathrm{exit}}=1$) \emph{and} it is safe to switch back. For the latter condition, it is important to carefully avoid problems that may arise from a possible periodic switching behavior between the two planners.

\subsubsection{Infeasibility Analysis}\label{Sec:InfeasibilityAnalysis}

The desirable outcome of the Primary MPC is a (local) minimum with an acceptable tolerance ($q_{\mathrm{exit}} = 1$). The SOTIF analysis in Section \ref{Sec:SOTIFAnalysis}, however, has revealed multiple scenarios where this may not be the case. Then a new module called the \emph{Infeasibility Analysis} is activated. Its logic is designed as follows:

If the Primary MPC has not converged to such a minimum and returns only a feasible trajectory ($q_{\mathrm{exit}} = 2$), the sub-optimal solution is acceptable on an exceptional basis. Because the sub-optimal solution satisfies also the terminal constraints, the recursive feasibility of the trajectory planning is not jeopardized. However, situations where the solver detects the definite infeasibility of the OCP ($q_{\mathrm{exit}} = 3$) or where it is unable to find a feasible solution ($q_{\mathrm{exit}} = 4$) require further attention. Here a switching logic, as presented in Table \ref{Tab:SwitchLogic}, is proposed as a basis for the Infeasibility Analysis.

\begin{table}[h]
	\renewcommand{\arraystretch}{1.5}
    \centering
    \begin{tabular}{|p{0.3cm}|p{2.0cm}|p{1.55cm}|p{1.3cm}|p{1.3cm}|p{1.3cm}|p{1.3cm}|}
    \cline{3-7}
    \multicolumn{2}{c|}{$\,$} & \multicolumn{5}{c|}{\textbf{Primary MPC (exit flag)}}\\ \cline{3-7}
    \multicolumn{2}{c|}{$\,$} & $q_{\mathrm{exit}}^{\mathrm{(P)}} = 1$\newline at\,$k=0,1$
    & $q_{\mathrm{exit}}^{\mathrm{(P)}} = 2$\newline \,at $k=0$
    & $q_{\mathrm{exit}}^{\mathrm{(P)}} = 2$\newline \,at $k=1$
    & $q_{\mathrm{exit}}^{\mathrm{(P)}} = 3$\newline \,at $k=1$
    & $q_{\mathrm{exit}}^{\mathrm{(P)}} = 4$\newline \,at $k=1$\\
    \hline\hline
    \multirow{5}{0.3cm}{\rotatebox{90}{\parbox[t]{4.8cm}{\textbf{Backup MPC (exit flag)}}}}
    & $q_{\mathrm{exit}}^{\mathrm{(B)}} = 1$\newline at\;$k=1$
    & \cellcolor{LiteGreen}\hspace*{0.7cm}P & \cellcolor{LiteGreen}\hspace*{0.55cm}P & \cellcolor{LiteYellow}\hspace*{0.55cm}B\newline at $k=1$ &\cellcolor{LiteYellow}\hspace*{0.55cm}B\newline at $k=1$ &\cellcolor{LiteYellow}\hspace*{0.55cm}B\newline at $k=1$\\ \cline{2-7}
    & $q_{\mathrm{exit}}^{\mathrm{(B)}} = 2$\newline at\;$k=1$
    & \cellcolor{LiteGreen}\hspace*{0.7cm}P & \cellcolor{LiteGreen}\hspace*{0.55cm}P & \cellcolor{LiteGreen}\hspace*{0.55cm}P &\cellcolor{LiteYellow}\hspace*{0.55cm}B\newline at $k=1$ &\cellcolor{LiteYellow}\hspace*{0.55cm}B\newline at $k=1$\\ \cline{2-7}
    & $q_{\mathrm{exit}}^{\mathrm{(B)}} = 3$\newline at\;$k=1,2$
    & \cellcolor{LiteGreen}\hspace*{0.7cm}P & \cellcolor{LiteGreen}\hspace*{0.55cm}P & \cellcolor{LiteGreen}\hspace*{0.55cm}P &\cellcolor{LiteRed}\hspace*{0.4cm}EO\newline at $k=1$ &\cellcolor{LiteBlue}\hspace*{0.4cm}SP\\ \cline{2-7}
    & $q_{\mathrm{exit}}^{\mathrm{(B)}} = 4$\newline at\;$k=1,2$
    & \cellcolor{LiteGreen}\hspace*{0.7cm}P & \cellcolor{LiteGreen}\hspace*{0.55cm}P & \cellcolor{LiteGreen}\hspace*{0.55cm}P &\cellcolor{LiteBlue}\hspace*{0.4cm}SP &\cellcolor{LiteBlue}\hspace*{0.4cm}SP\\ \cline{2-7}
    & $q_{\mathrm{exit}}^{\mathrm{(B)}} = 3,4$\newline at\;$k=3$
    & \cellcolor{LiteGreen}\hspace*{0.7cm}P & \cellcolor{LiteGreen}\hspace*{0.55cm}P & \cellcolor{LiteGreen}\hspace*{0.55cm}P &\cellcolor{LiteRed}\hspace*{0.4cm}EO\newline at $k=3$ &\cellcolor{LiteRed}\hspace*{0.4cm}EO\newline at $k=3$\\ \hline
    \end{tabular}
    \vspace*{0.1cm}\\
    \footnotesize \textbf{Legend:} P: continue with Primary MPC solution; B: switch to Backup MPC;\\ EO: switch to Emergency Operation; SP: use stored Primary MPC solution
    \caption{Partial illustration of the switching logic when the Primary MPC is active.\label{Tab:SwitchLogic}}
\end{table}

Before even activating the switching logic, however, the Infeasibility Analysis filters a few special situations that will obviously or possibly lead to an infeasibility of the Primary MPC (\FIref{FI:Infeasible}). This includes 
\begin{itemize}
    \item (\TCref{TC:AOUCS05}) a leading TgtV driving in reverse, while the EgoV is at rest or slowly moving forward; 
    \item (\TCref{TC:ASA04}) at higher speeds, the cut-in of a faster TgtV in front of the EgoV leading to a violation of the safety distance in the present time step.
\end{itemize}

For (\TCref{TC:AOUCS05}), a special planning mode shall be designed to check if (a) the EgoV is driving with a speed less than or equal to some threshold $v_{\mathrm{low}}$ (e.g., $v_{\mathrm{low}}=5\,\frac{\mathrm{km}}{\mathrm{h}}$) and (b) the minimum safety distance to the leading TgtV is already violated at the current time step. If this is the case, a stopping trajectory is automatically issued that transitions the EgoV to, or keeps it in, a complete standstill. When the Primary MPC becomes feasible again, the EgoV resumes its normal operation and moves forward. 

For (\TCref{TC:ASA04}), the situation can be resolved by an additional logic that relaxes the minimum safety distance constraint to a leading TgtV in the Primary MPC, if it has a higher velocity than the EgoV. Note that, even if the safety distance is not violated, the cut-in of a faster TgtV (\TCref{TC:ASA04}) may still lead to an overly conservative braking maneuver on behalf of the EgoV (\FIref{FI:AOUCS04}c). This is a separate issue that has to be resolved by a different measure.

In situations with a slippery road surface (\TCref{TC:ACTC09}) or a sudden drop of the friction value $\mu$ within a time interval of $t_d$ for more than $\mu_{d}$ (\TCref{TC:AOUCS07}), in a first step the linear tire regime of the LDBM in (\ref{Equ:MSSP}b) shall be exchanged with a (nonlinear) Pacejka tire model. This measure is taken pre-emptively in order to avoid a lack of drivability of the computed trajectory (\FIref{FI:Drivable}a). The Pacejka tire model is not used as the standard tire model of the Primary MPC because of its additional computational complexity.

%In the case of $q_{\mathrm{exit}}=3$, the Primary MPC-TP detects that there exists no trajectory for the given set of constraints (\FIref{FI:Infeasibility}), by a certificate of infeasibility. As a first step, all constraints that are not strictly necessary shall be relaxed, i.e., the traffic rule and comfort constraints $\mathbb{X}^{(\mathrm{t})}_{t}$ are removed from (\ref{Equ:MSSP}f). Then the corresponding \emph{Relaxed OCP} (ROCP) is re-solved. If the ROCP has a feasible solution, the resulting trajectory is used and the MPC-TP1 continues as normal. If the ROCP again has no feasible solution, 

According to Table \ref{Tab:SwitchLogic}, if the Primary MPC does not produce a feasible solution ($q_{\mathrm{exit}}^{\mathrm{(P)}} = 3,4$), the Backup MPC is activated already after 2 steps. It should be remarked that this could be extended, if the reaction horizon were increased beyond $N_{\mathrm{r}}=4$, which is used in this paper. This may allow for further attempts to modify or update the constraints of the OCP in an acceptable way, similar to the steps described above. Or additional attempts may be taken to modify the algorithm or the solver settings in order to achieve an optimal, or at least a feasible solution, of the OCP.

%In particular, the Primary MPC may attempt to compute a new trajectory from a current state measurement (instead of the predicted state), while again relaxing the traffic rule and comfort constraints $\mathbb{X}^{(\mathrm{t})}_{t}$. If all attempts fail ($q_{\mathrm{exit}}=3,4$) until the maximum computation time $\bar{t}_{\mathrm{c}}$ is reached, the Backup MPC is checked for an optimal or feasible solution ($q_{\mathrm{exit}}=1,2$), which is then used. If the Backup MPC also fails ($q_{\mathrm{exit}}=3,4$), the emergency operation of the encompassing system is activated, which transfers the EgoV to a safe state, e.g. by braking and driving to one side of the lane.

%In the case where $q_{\mathrm{exit}}=4$, the solver of the Primary MPC is unable to find a feasible solution (\FIref{FI:NoSolution}), but without a certificate of infeasibility. Since software and hardware faults fall into the scope of FS, this must be due to unpredictable numerical problems (e.g., due to floating point issues or overflow). To the best of the authors' knowledge, there are no known specific TCs for this case. The procedure to handle it is the same as for the case $q_{\mathrm{exit}}=3$, which is described above.

\subsubsection{MPC-TP Driving Modes}

To adequately adapt to specific traffic or environmental situations, the MPC-TP needs to introduce a switchable behavior. {\Spec\label{Spec:SpecialDrivingModes}}*) In particular, (at least)the specialized driving modes
listed in Table \ref{Tab:DrivingModes} shall be introduced.

\begin{table}
    \vspace*{-0.5cm}
    \small
    \renewcommand{\arraystretch}{1.5}
    \centering
      \begin{tabular}{|p{2.7cm} | p{5.8cm} | p{2.2cm}|}
        \hline
         \textbf{Mode} & \textbf{Description} & \textbf{Addressed\linebreak FIs / TCs}\\
         \hline\hline
         Obstacle \mbox{Avoidance} & Situations where obstacles require an emergency evasive maneuver, e.g., within the current lane, by driving in between two lanes or by using the shoulder &  (\TCref{TC:AOUCS02}) i.c.w.\ (\FIref{FI:AOUCS02})\\
         \hline
         Rescue Alley & Situations that require the formation of a rescue alley, e.g., during a congestion or in the presence of an emergency vehicle&  (\TCref{TC:AOUCS01}) i.c.w.\ (\FIref{FI:AOUCS01}) \\
         \hline
         Oversized Vehicle / Heavy Transport & Situations involving an oversized vehicle or a heavy duty transport in front of the EgoV, require careful driving, driving to one side of a lane or between two lanes, and passing restrictions & (\TCref{TC:EAL06}) i.c.w.\ (\FIref{FI:InsufficientTime}), (\FIref{FI:Infeasible}), or (\FIref{FI:DangerousManeuver})\\
         \hline
         Wrong Way Driver & Potential encounter with a wrong-way driver, requiring slow driving on the outer right-hand lane without overtaking and keeping a generous safety distance&
         (\TCref{TC:AOUCS06}) i.c.w.\ (\FIref{FI:InsufficientTime}), (\FIref{FI:Infeasible}), or (\FIref{FI:DangerousManeuver})\\
         \hline
         Trafficable Hard Shoulder & Situations where the freeway shoulder is open to traffic & (\TCref{TC:EAL10}) i.c.w.\ (\FIref{FI:InsufficientTime}), (\FIref{FI:Infeasible}), or (\FIref{FI:DangerousManeuver}) \\
         \hline
         Obstacle on Hard Shoulder & Situations where the freeway shoulder is blocked by an obstacle (e.g., by a damaged vehicle) or occupied with persons,  requiring careful driving and ideally using the left lanes&(\TCref{TC:EAL07}) i.c.w.\ (\FIref{FI:InsufficientTime}), (\FIref{FI:Infeasible}), or (\FIref{FI:DangerousManeuver}) \\
         \hline
         Defensive Driving & General defensive driving (reduced speed, no overtaking, no lane changing, increase safety distances), e.g., due to dropped cargo or large animals crossing the freeway & (\TCref{TC:AFD01}c)  i.c.w.\ (\FIref{FI:InsufficientTime}), (\FIref{FI:Infeasible}), or (\FIref{FI:DangerousManeuver})\\
         \hline
         Normal Driving & All other driving situations & N/A\\
         \hline
    \end{tabular}
    \normalsize
    \caption{Specialized driving modes (i.c.w. = `in combination with').\label{Tab:DrivingModes}}
\end{table}

The proper driving mode needs to be selected by the overarching system and provided to the primary as well as the back-up MPC-TP as additional mode inputs. The MPC-TPs can then adjust their behaviors considering the detected situation. This goes hand in hand with integration requirements to detect the various driving modes.

\subsection{Refined ODD}\label{Sec:RefinedODD}

\Specref{Spec:ODD}i*) To address issues, where the prediction horizon $N_{\mathrm{p}}$ of the EgoV is not large enough for the EgoV to come to a full stop if needed (cf. \FIref{FI:AOUCS03} in combination with (\TCref{TC:AOUCS03}), we limit the maximum permissible speed in the ODD: 
\begin{equation}\label{Equ:MaxVel} v_{\max}=N_{\mathrm{p}}\,a_{\mathrm{lon},\min}\,t_{\mathrm{s}}\;.
\end{equation}
As the prediction horizon $N_{\mathrm{p}}$ is static,  $v_{\max}$ must be chosen in accordance with the maximum deceleration $a_{\mathrm{lon},\min}$.
$a_{\mathrm{lon},\min}\leq 0$ is the worst-case (highest) deceleration, depending on current road conditions. It depends on the worst-case values of other ODD parameters, in particular the friction value $\mu$.  
%Alternatively, the MPC-TP could include a more complex terminal condition for recursive feasibility, as demanded in (\Specref{Spec:MPC}e), in order to enhance its performance.

\Specref{Spec:ODD}o*) To cope with (FI005) in combination with (TC015c) the ODD will be limited to freeway sections with a game fence. 

(\Specref{Spec:ODD}f*,h*) To mitigate (\FIref{FI:Infeasible}) in combination with (\TCref{TC:AOUCS07}), in particular with (\TCref{TC:AOUCS07}b), we propose to limit the minimum ambient temperature to $+4^{\circ}\mathrm{C}$  (\Specref{Spec:ODD}f*) and the permissible minimum road friction value (\Specref{Spec:ODD}h*). This will be combined with an integration requirement to conservatively estimate the friction value ({\IRref{IR:FrictionEstimation}}), cf.\ Section \ref{Sec:IntegrationReqs}.  

The refined ODD is provided in in Table \ref{Tab:RefinedODDDescription}. 

\begin{table}
    \vspace*{-0.5cm}
\small
\renewcommand{\arraystretch}{1.5}
\begin{tabular}{| p{3.7cm} | p{6cm} | p{1.0cm} |}
\hline 
  \multicolumn{2}{|l|}{\textbf{Physical Infrastructure}} & \textbf{Identi- fier}\\
\hline\hline
  roadway types: & German divided freeway (Autobahn) & (\Specref{Spec:ODD}a)  \\
  \hline 
  roadway surfaces: & asphalt or concrete slabs & (\Specref{Spec:ODD}b) \\
  \hline 
  roadway geometry: &  straightways or curves with bank angles $\leq8^{\circ}$ (max.\ value permitted for German roads) & (\Specref{Spec:ODD}c) \\
  \hline \hline 
  \multicolumn{3}{|l|}{\textbf{Objects}} \\
\hline\hline
  roadway users - vehicles: & all vehicles drive forward only & (\Specref{Spec:ODD}d)  \\
  \hline 
  non-roadway users - obstacles/objects: & no pedestrians or bicycles & (\Specref{Spec:ODD}e)\\
  \hline \hline  

 \hline 
  \multicolumn{3}{|l|}{\textbf{Operational Constraints}} \\
\hline\hline
  Speed limits - max. speed limit: & \textcolor{MedBlue}{$v_{\max}=N\,a_{\mathrm{lon},\min}\,t_{\mathrm{s}}$} & \textcolor{MedBlue}{(\Specref{Spec:ODD}i*)}  \\
  \hline \hline  
  
  \multicolumn{3}{|l|}{\textbf{Environmental Constraints}} \\
\hline \hline 
  weather - temperature: & ambient temperature within $[+4^{\circ}\mathrm{C}, +50^{\circ}\mathrm{C}]$ & (\Specref{Spec:ODD}f*) \\
  \hline 
  weather - wind: & wind speed below $8\,\mathrm{bft}$ ($\leq 74\,\frac{\mathrm{km}}{\mathrm{h}}$) & (\Specref{Spec:ODD}g) \\
 
  \hline 
  weather-induced roadway conditions - friction value: & \textcolor{MedBlue}{ $\mu >\mu_{\min}=0.15$} & \textcolor{MedBlue}{(\Specref{Spec:ODD}h*)} \\
  \hline
 \hline
  \multicolumn{3}{|l|}{\textbf{\textcolor{MedBlue}{Zones}}} \\
\hline\hline
  \textcolor{MedBlue}{Geo-fencing: }& \textcolor{MedBlue}{Game fence needs to be present} & \textcolor{MedBlue}{(\Specref{Spec:ODD}o*)} \\
    \hline  
  
\end{tabular}
\normalsize
\caption{Refined ODD. \textcolor{MedBlue}{(Changes to the original ODD are shown in blue.)}\label{Tab:RefinedODDDescription}}
\end{table}

\subsection{Integration Requirements}\label{Sec:IntegrationReqs}

A key concept to facilitate the out-of-context development of SOTIF-related elements are assumptions on the encompassing vehicle control system. As these need to be ensured during the integration of the SOTIF-EooC into the ADS, they shall be designated as integration requirements. The new integration requirements of the SOTIF analysis are listed in Table \ref{Tab:IntegrationRequirements}.

%A1: Surrounding vehicles either have the right-of-way, and are thus considered as obstacles, or they have to yield, and can thus be ignored. However, there are situations where the right-of-way is unclear, such as during a lane merging lane ending (\TCref{TC:AOUCS04}a). Without further measures, this may lead to an overly conservative driving behavior (\FIref{FI:AOUCS04}a), or even a complete deadlock for the EgoV(\FIref{FI:AOUCS04}b). This problem requires further analysis and has to be resolved in combination with the Tactical Planner.

\begin{table}[H]
    \vspace*{-0.1cm}
    \small
    \renewcommand{\arraystretch}{1.5}
    \centering
      \begin{tabular}{|p{1.5cm} | p{7.5cm} | p{1.7cm}|}
        \hline
         \textbf{Identifier} & \textbf{Description} & \textbf{Addressed\linebreak FIs / TCs}\\
         \hline\hline
         (\IRref{IR:RescueAlley}a) & Congestion situations requiring the formation of a rescue alley shall be detected by the encompassing system. & \multirow{2}{1.8cm}{(\FIref{FI:AOUCS01}) i.c.w.\ (\TCref{TC:AOUCS01})}\\\cline{1-2}
         (\IRref{IR:RescueAlley}b) & In case of a detected congestion situation that requires the formation of a rescue alley, the encompassing system shall request the mode 'rescue alley' from the MPC. & \\\hline
         (\IRref{IR:ObstAvoidance}a) & Obstacle avoidance situations shall be detected by the encompassing system. & \multirow{2}{1.8cm}{(\FIref{FI:AOUCS02}) i.c.w.\ (\TCref{TC:AOUCS02})}\\\cline{1-2}
         (\IRref{IR:RescueAlley}b) & In case of a detected obstacle avoidance situation, the encompassing system shall request the mode 'obstacle avoidance' from the MPC, in which the lane keeping constraints are relaxed, but the collision constraints remain intact. & \\\hline
         (\IRref{IR:WrongWayDriver}) & Situations with wrong way drivers shall be detected and handled by the encompassing system utilizing a dedicated emergency planner that is independent of the MPC-TP. & (\FIref{FI:Infeasible}) i.c.w.\ (\TCref{TC:AOUCS06})\\\hline
         (\IRref{IR:FrictionEstimation}a) &The friction value $\mu$ shall be conservatively estimated for a prediction time $T_{\mathrm{p}}=6\,\mathrm{s}$. & \multirow{2}{1.8cm}{(\FIref{FI:Infeasible}) i.c.w.\ (\TCref{TC:AOUCS07})} \\\cline{1-2} ({\IRref{IR:FrictionEstimation}b}) &It shall be ensured that the actual friction value does not deviate from the actual friction value by more than a tolerance $\Delta\mu$. This will be combined with ODD restriction (\Specref{Spec:ODD}h). & \\\hline
         (\IRref{IR:EmergencyOperationRequest}) & Situations where the MPC-TP requests an emergency operation shall be handled by the encompassing system utilizing a dedicated emergency planner that is independent of the MPC-TP. & \\\hline
         (\IRref{IR:IrregularRoadSurface}) & Situations involving irregular road surfaces shall be detected and handled by the encompassing system utilizing the Tactical Planner. &
         (\FIref{FI:Infeasible}) i.c.w.\ (\TCref{TC:AOUCS08})\\\hline
    \end{tabular}
    \normalsize
    \caption{Overview of integration requirements (i.c.w. = `in combination with').\label{Tab:IntegrationRequirements}}
\end{table}

Cases where neither the Primary MPC nor the Back-Up MPC is able to generate a permissible trajectory shall be handled by the encompassing system as follows:\\
(\IRref{IR:EmergencyOperationRequest}) Situations where the MPC-TP requests an emergency operation shall be handled by the encompassing system utilizing a dedicated emergency planner that is independent of the MPC-TP.

Situations involving irregular road surfaces (\TCref{TC:AOUCS08}) where the MPC-TP would be unable to generate a feasible (collision-free) trajectory (\FIref{FI:Infeasible}) shall be identified and handled by the encompassing system. Here, the Tactical Planner needs to decide what to in such cases, e.g., depending on the size of the stones or potholes and provide an appropriate tactical decision to the MPC-TP or handle the situation without the MPC-TP.
This leads to integration requirement (\IRref{IR:IrregularRoadSurface}): Situations involving irregular road surfaces shall be detected and handled by the encompassing system utilizing the Tactical Planner.

\subsection{Situations Requiring Further Analysis}\label{Sec:FurtherAnalysis}

(Further analysis required) The OCP is based on the LDBM, assuming a linear tire model and no roll motion, which may represent a poor approximation of a real vehicle if these assumptions are violated (\FIref{FI:AAOD01}). The non-negligible roll motion may be violated by an extremely asymmetric load (\TCref{TC:AAOD01}a), a highly dynamic cornering maneuver (\TCref{TC:AAOD01}b), or a high position of the CoG (\TCref{TC:AAOD01}c). A detailed simulation analysis is required to examine whether the LDBM is sufficient for the targeted use cases (especially with respect to \SRref{SR:Drivable}), or whether a two-track model is required instead.

Situations encompassing side wind gusts (\TCref{TC:ASA01}) are deemed as adequately covered by the wind limitation in the ODD (\Specref{Spec:ODD}g).

Potentially unknown traffic scenarios or traffic signs, e.g., due to deviating regulations in other European countries, are potentially problematic as well (\TCref{TC:EAL15}). However, the existing ODD restriction limits the permissible roadway types to the German Autobahn (\Specref{Spec:ODD}a) seems sufficient to mitigate this concern.

\subsection{Interaction-aware MPC}\label{Sec:InteractiveMPC}

Particular combinations of TCs and FIs cannot be adressed without extending the architecture of the encompassing vehicle, e.g. by adding additional senors and/or capabilities. An example is the combination of (\TCref{TC:AOUCS04}a) and (\FIref{FI:AOUCS04}).
As per our analysis the OCP in \eqref{Equ:MSSP} may lead to an overly conservative behavior of the EgoV (\FIref{FI:AOUCS04}), possibly leading to poor progress on the road (\FIref{FI:AOUCS04}a), or even a deadlock (\FIref{FI:AOUCS04}b). The only safe way to reduce this conservatively is to internalize the relevant TgtVs in the MPC predictions and to reflect the interactions between the TgtVs and the EgoV by appropriate behavioral models. Interaction-aware MPC approaches, however, are the subject of ongoing and future research \cite{NairEtAl:2022,BencioliniEtAl:2023,ChenEtAl:2022,OliveiraEtAl:2023} and are beyond the scope of this paper.

\section{Conclusion}\label{Sec:Conclusion}

In this paper, multiple aspects of the general SOTIF framework provided in ISO\,21448 have been applied to a generic MPC-based trajectory planner, which is part of an ADS. The generic MPC-TP is treated as an SOTIF-EooC, elaborating this concept that is only sketched in the SOTIF standard. In a first step, the initial concept of a generic MPC-TP has been analyzed in order to identify relevant FIs and corresponding TCs. To this end, the analysis techniques that are described in ISO\,21448 only in very general terms have to be interpreted and applied to the case study. In a second step, the resulting comprehensive list of relevant FIs and TCs informed an SOTIF-optimized concept of the MPC-TP. The revised safety architecture includes, in particular, an infeasibility analysis as a component and a back-up trajectory planner, a switchable behavior of the MPC-TP to support special driving modes, and adaptations to the reference trajectory.  

An important aspect of the results is the identification of important research topics on MPC for path planning in the context of automated and autonomous driving, from a safety perspective:
\begin{itemize}
\item[(i)] interaction-awareness with other (human) traffic participants;
\item[(ii)] ability to avoid obstacles and drive safely even at low friction values;
\item[(iii)] design of a recursively feasible MPC-based path planner with a reaction horizon;
\item[(iv)] string stability in a chain of vehicles;
\item[(v)] examination of a safe switching logic between a Primary MPC and a Backup MPC. 
\end{itemize}

In future work, an extensive simulation study would be beneficial for a more detailed understanding of the potential shortcomings of the MPC approach itself, or the respective vehicle model to be used for its predictions in extreme driving situations.

\section*{Acknowledgment}\label{ACKNOWLEDGMENT}

The research leading to these results was funded by the German Federal Ministry for Economic Affairs and Climate Action (BMWK) through the project EEMotion.

%Our work identified multiple areas for future research on realizing the SOTIF for real-world applications including: (i) further elaboration of the SOTIF-EooC concept;
%(ii) Elaboration of detailed methodological guidance for the analysis techniques for the identification of FIs and TIs; and (iii) analysis of the end-to-end timing behavior of SOTIF-related systems-

\newpage

\begin{landscape}
    \begin{textblock*}{10cm}(4.5cm,2cm)
        \section*{Appendix A: Overview of TCs and FIs}\label{Sec:Appendix}
    \end{textblock*}
    \small
    \begin{longtable}[c]{|p{6cm}|p{6cm}|p{7cm}|}
    \hline
    \textbf{TC} & \textbf{Triggered FI} & \textbf{Notes}\\
\hline\hline

({\TC\label{TC:AOUCS01}}) traffic congestion
& 
({\FI\label{FI:AOUCS01}}) no capability to form a rescue alley for police or ambulances\newline
({\FI\label{FI:DangerousManeuver}}) generation of trajectories representing unusual, illegal, counter-intuitive, or dangerous maneuvers
& 
(refined specification/design) introduce switchable behavior of the MPC (driving mode: Rescue Alley)
\newline
(integration requirement) congestion situations shall be detected and handled by the encompassing system ({\IR\label{IR:RescueAlley}}a,b)\\
\hline

({\TC\label{TC:AOUCS02}}) obstacle avoidance scenario 
& 
({\FI\label{FI:AOUCS02}}) no capability to carry out emergency maneuvers without a proper lane change, driving in between two lanes, or by using the shoulder\newline
(\FIref{FI:DangerousManeuver}) generation of trajectories representing unusual, illegal, counter-intuitive, or dangerous maneuvers
& 
(refined specification/design) introduce switchable behavior of the MPC (driving mode: Obstacle Avoidance)\newline
(integration requirement) obstacle avoidance situations shall be detected and handled by the encompassing system ({\IR\label{IR:ObstAvoidance}}a,b)\\
\hline

({\TC\label{TC:AOUCS03}}) blocking obstacle shortly beyond the prediction horizon $N_{\mathrm{p}}$ of the MPC-TP and no obstacle avoidance is possible 
& 
({\FI\label{FI:AOUCS03}}) prediction horizon $N_{\mathrm{p}}$ of the MPC-TP is too short to come to a full stop within it
& 
(ODD restriction) limit permissible max. speed to
$v_{\max}=\left|N_{\mathrm{p}}\,a_{\mathrm{lon},\min}\,t_{\mathrm{s}}\right|$ (\Specref{Spec:ODD}i*)\\
\hline

({\TC\label{TC:AOUCS04}}a) right-of-way is unclear: at the end of two merging lanes \newline
(\TCref{TC:AOUCS04}b) right-of-way is unclear: special situations during general lane changes 
&
({\FI\label{FI:AOUCS04}}a) overly conservative behavior of the MPC-TP: inability to ensure sufficient progress on the road\newline
(\FIref{FI:AOUCS04}b) overly conservative behavior of the MPC-TP: inability to avoid deadlocks in certain situations \newline
(\FIref{FI:DangerousManeuver}) generation of trajectories representing unusual, illegal, counter-intuitive, or dangerous maneuvers
& 
(Interaction-aware MPC) interaction-aware MPC approach, in coordination with the Tactical Planner (\Specref{Spec:InteractionAwareMPC}*)\\
\hline

({\TC\label{TC:AOUCS05}}) Leading TgtV driving in reverse, in combination with a stopping or slowly moving EgoV 
& 
({\FI\label{FI:Infeasible}}) inability to generate a feasible (collision-free) trajectory in certain situations 
& 
(refined specification/design) infeasibility analysis: selection of stopping trajectory in situations with low speed and safety distance violated\\
\hline

({\TC\label{TC:AOUCS06}}) wrong way driver 
& 
(\FIref{FI:Infeasible}) inability to generate a feasible (collision-free) trajectory in certain situations \newline
(\FIref{FI:DangerousManeuver}) generation of trajectories representing unusual, illegal, counter-intuitive, or dangerous maneuvers
& 
(refined specification/design) introduce switchable behavior of the MPC (driving mode: Wrong Way Driver)\newline
(integration requirement) if infeasibility still occurs, switch to emergency operation of the encompassing system ({\IR\label{IR:EmergencyOperationRequest}})\newline
(integration requirement) situations with wrong way drivers shall be detected and handled by dedicated mode of the encompassing system ({\IR\label{IR:WrongWayDriver}})\\
\hline

({\TC\label{TC:AOUCS07}}a) sudden drop of friction coefficient: water (hydroplaning) \newline
(\TCref{TC:AOUCS07}b) sudden drop of friction coefficient: snow / ice \newline
(\TCref{TC:AOUCS07}c) sudden drop of friction coefficient: oil spillage 
& 
(\FIref{FI:Infeasible}) inability to generate a feasible (collision-free) trajectory in certain situations 
&
(ODD restriction) limit permissible minimum temperature to exclude snowy/icy conditions (\Specref{Spec:ODD}f*) 
\newline
(ODD restriction) limit permissible minimum friction value (\Specref{Spec:ODD}h*)\newline
(integration requirement) conservative estimation of the friction value ({\IR\label{IR:FrictionEstimation}a})\newline
(refined specification/design) avoid small areas filled with water by evasive maneuvers within the lane boundaries\\
%for the prediction time $T_{\mathrm{p}}=6\,\mathrm{s}$ (cf.\Specref{Spec:EnvironmentFusion}a)\\
\hline

(\TCref{TC:AOUCS07}a) sudden drop of friction coefficient: water (hydroplaning)\newline
(\TCref{TC:AOUCS07}b) sudden drop of friction coefficient: snow, ice \newline
(\TCref{TC:AOUCS07}c) sudden drop of friction coefficient: oil spillage 
& 
({\FI\label{FI:Drivable}}a) inability to generate a drivable trajectory in certain situations: due to linear tire model in LDBM
& (refined specification/design) infeasibility analysis: temporarily relax the linear tire regime if the friction value drops by more than $\mu_{\mathrm{d}}$ within $t_{\mathrm{d}}$\\
\hline

({\TC\label{TC:AOUCS08}}a) irregular road surface: gravel or stones\newline
(\TCref{TC:AOUCS08}b) irregular road surface: potholes\newline
(\TCref{TC:AOUCS08}c) irregular road surface: road bumps
 & 
 (\FIref{FI:Drivable}b) inability to generate a drivable trajectory in certain situations: due to irregular road surface
 &
(integration requirement) situations with irregular road surface shall be detected and handled by the encompassing system in combination with the Tactical Planner ({\IR\label{IR:IrregularRoadSurface}})\\
\hline

({\TC\label{TC:ASA01}}a) side wind gusts: bridge deck\newline
(\TCref{TC:ASA02}a) irregular lane markings: suddenly disappearing lane markings\newline
({\TC\label{TC:ASA02}}b) irregular lane markings: dirty lane markings\newline
(\TCref{TC:ASA02}c) irregular lane markings: broken lane markings\newline
(\TCref{TC:ASA02}d) irregular lane markings: covered lane markings 
& 
({\FI\label{FI:ASA01}}) receding-horizon planning of the MPC-TP in combination with tracking controller could lead to oscillations, or even an unstable behavior of the EgoV 
& 
(refined specification/design) MPC-TP to keep the first $N_{\mathrm{r}}=4$ steps (reaction horizon) of the trajectory unchanged (\Specref{Spec:TrajectoryBody}g*) \\
\hline

(\TCref{TC:AOUCS04}) right-of-way is unclear\newline
({\TC\label{TC:ASA05}}) other traffic participants not strictly adhering to the traffic rules 
& (\FIref{FI:AOUCS04}a) overly conservative behavior of the MPC-TP: inability to ensure sufficient progress on the road\newline
(\FIref{FI:AOUCS04}b) overly conservative behavior of the MPC-TP: inability to avoid deadlocks in certain situations\newline
(\FIref{FI:DangerousManeuver}) generation of trajectories representing unusual, illegal, counter-intuitive, or dangerous maneuvers
&
(Interaction-aware MPC) interaction-aware MPC approach, in coordination with the Tactical Planner (\Specref{Spec:InteractionAwareMPC}*)\\
\hline

\multirow{2}{6cm}{({\TC\label{TC:ASA04}}) cut-in of a faster TgtV in front of the EgoV}
& 
(\FIref{FI:AOUCS04}c) overly conservative behavior of the MPC-TP: overly conservative braking in cut-in scenarios 
& 
(refined specification/design) tuning of MPC-TP such that it avoids overly conservative braking to a leading TgtV with a positive relative velocity\\
\cline{2-3}
& 
(\FIref{FI:Infeasible}) inability to generate a feasible (collision-free) trajectory in certain situations
& 
(refined specification/design) infeasibility analysis: if TgtV cuts into the EgoV's lane and is faster than EgoV, then temporarily relax linear minimum distance to this TgtV\\
\hline

({\TC\label{TC:ASA06}}) trailing vehicle driving closely behind the EgoV in the same lane, not keeping a proper safety distance 
& ({\FI\label{FI:ASA03}}) inability of the MPC-TP to consider trailing traffic in the lane of the EgoV, where a sudden braking maneuver of the EgoV may lead to a collision 
&
(refined specification/design) MPC-TP to brake only when strictly necessary and only to an appropriate extent (\Specref{Spec:SuddenBraking}a*) \newline 
(refined specification/design) MPC-TP to account for distance to trailing vehicle by means of tuning the weight of $\ell_{\mathrm{o}}(\cdot,\cdot)$ in the cost function (\Specref{Spec:SuddenBraking}b*) \\
\hline

\multirow{2}{6cm}{({\TC\label{TC:AFD01}}a) suddenly appearing obstacle: previously shadowed traffic participant\newline
(\TCref{TC:AFD01}b) suddenly appearing obstacle: object dropping from a bridge or truck\newline
(\TCref{TC:AFD01}c) suddenly appearing obstacle: large animal (e.g., wild boar) crossing the freeway}
& 
(\FIref{FI:Infeasible}) inability to compute a feasible (collision-free) trajectory in certain situations \newline
({\FI\label{FI:InsufficientTime}}) insufficient time to solve the OCP, in order to compute a new trajectory
&
(refined specification/design) infeasibility analysis\newline
(refined specification/design) potential extension of computation time to multiple time steps based on the concept of a reaction horizon\\
\cline{2-3}
&
(\FIref{FI:DangerousManeuver}) generation of trajectories representing unusual, illegal, counter-intuitive, or dangerous maneuvers
& 
(ODD restriction) limit ODD to freeway sections with a game fence (\Specref{Spec:ODD}o*) \newline
(refined specification/design) introduce switchable behavior of the MPC (driving mode: Defensive Driving)\\
\hline

({\TC\label{TC:AFD02}}a) sudden traffic rule change: start of a speed limit area\newline
(\TCref{TC:AFD02}b) sudden traffic rule change: start of a no passing zone 
& 
(\FIref{FI:Infeasible}) inability to generate a feasible (collision-free) trajectory in certain situations\newline
(\FIref{FI:InsufficientTime}) insufficient time to solve the OCP, in order to compute a new trajectory 
&
(ODD restriction) limit permissible max. speed to
$v_{\max}=\left|N_{\mathrm{p}}\,a_{\mathrm{lon},\min}\,t_{\mathrm{s}}\right|$ (\Specref{Spec:ODD}i*)\newline
(refined specification/design) infeasibility analysis\newline
(refined specification/design) potential extension of computation time to multiple time steps based on the concept of a reaction horizon
\\
\hline

({\TC\label{TC:AAOD01}}a) LDBM assumption violation: asymmetric load\newline
(\TCref{TC:AAOD01}b) LDBM assumption violation: dynamic cornering\newline
(\TCref{TC:AAOD01}c) LDBM assumption violation: high position of the CoG 
& 
({\FI\label{FI:AAOD01}})  poor approximation of real vehicle behavior by the LDBM in the OCP
& 
(further analysis required) experimental analysis: evaluate if MPC-TP behaves safely in these situations; otherwise consider usage of a more complex model (e.g., a two track model) in the OCP\\
\hline

({\TC\label{TC:StringStability}}) chain of vehicles, EgoV in the middle, first vehicle brakes suddenly 
& 
({\FI\label{FI:StringStability}}) lack of string stability of MPC-TP in a chain of vehicles, potentially leading to collisions
& 
(refined specification/design) proper tuning of MPC-TP for string stability, to be verified in simulations and/or on-road tests\\
\hline

({\TC\label{TC:SlowLeadingTgtV}}) fast approach of a slowly driving TgtV in the front of the EgoV, where the TgtV cannot be avoided by a lateral maneuver

&
(\FIref{FI:Infeasible}) inability to generate a feasible (collision-free) trajectory in certain situations\newline
({\FI\label{FI:RecursiveFeasibility}}) lack of recursive feasibility 
& 
(refined specification/design) infeasibility analysis\newline
(refined specification/design) if collision is unavoidable, activation of a collision mitigation mode in the encompassing system\\
\hline

({\TC\label{TC:AAOD02}}) rapidly changing traffic situations 
%\newline(\TCref{TC:AFD01}) suddenly appearing obstacle \textcolor{red}{Eintrag gab es schon einmal weiter oben}
&
(\FIref{FI:InsufficientTime}) insufficient time to solve the OCP, in order to compute a new trajectory 
&
(refined specification/design) potential extension of computation time to multiple time steps based on the concept of a reaction horizon\\
\hline

\textit{no specific TCs known}
&
({\FI\label{FI:NoSolution}}) inability to generate a solution in certain situations at all 
&
(refined specification/design) infeasibility analysis
\\\hline

Pairs of extreme parameter values in some or all corner cases:\newline 
({\TC\label{TC:CommonTrigger1}}) max.\ side wind speed $v_{\mathrm{wind},\max} = 8\,\mathrm{bft}$ and max.\ bank angle $\beta_{\max} = 8^{\circ}$ \newline
({\TC\label{TC:CommonTrigger2}}) min.\ friction value $\mu_{\min}=0.15$ and min.\ tire cornering stiffness $C_{\alpha,\mathrm{f},\min}=60\,\frac{\mathrm{kN}}{\mathrm{rad}}$\newline 
({\TC\label{TC:CommonTrigger3}}a) max.\ front tire cornering stiffness $C_{\alpha,\mathrm{f},\max}=200\,\frac{\mathrm{kN}}{\mathrm{rad}}$ $\times$ and min.\ rear tire cornering stiffness $C_{\alpha,\mathrm{r},\min}=60\,\frac{\mathrm{kN}}{\mathrm{rad}}$\newline
(\TCref{TC:CommonTrigger3}b) min.\ front tire cornering stiffness $C_{\alpha,\mathrm{f},\min}=60\,\frac{\mathrm{kN}}{\mathrm{rad}}$ $\times$ and max.\ rear tire cornering stiffness $C_{\alpha,\mathrm{r},\max}=200\,\frac{\mathrm{kN}}{\mathrm{rad}}$\newline
...
&
(\FIref{FI:Drivable}a) inability to generate a drivable trajectory in certain situations: due to linear tire model in LDBM\newline
(\FIref{FI:AAOD01}) poor approximation of real vehicle behavior by the LDBM in the OCP
& 
(further analysis needed) experimental analysis for pairs of boundary values, conduct simulations to confirm/discard relevance of LDBM for MPC-TP; otherwise consider usage of a more complex model (e.g., a two track model) in the OCP\\
\hline

(\TCref{TC:ASA02}a) irregular lane markings: suddenly disappearing lane markings \newline
(\TCref{TC:ASA02}b) irregular lane markings: dirty lane markings\newline
(\TCref{TC:ASA02}c) irregular lane markings: broken lane markings\newline
(\TCref{TC:ASA02}d) irregular lane markings: covered lane markings 
& 
& 
(further analysis needed) common TCs, conduct simulations to confirm/discard relevance for MPC-TP \\
\hline

({\TC\label{TC:ACTC03}}a) unrecognizable lane markings: direct sunlight \newline
(\TCref{TC:ACTC03}b) unrecognizable lane markings: change of lighting conditions, e.g., when entering a tunnel 
& 
& 
(further analysis needed) common TCs, conduct simulations to confirm/discard relevance for MPC-TP \\
\hline

({\TC\label{TC:ACTC04}}a) insufficient visibility of lane markings: snow / ice  \newline
(\TCref{TC:ACTC04}b) insufficient visibility of lane markings: fog 
& 
& 
(further analysis needed) common TCs, conduct simulations to confirm/discard relevance for MPC-TP \\
\hline

({\TC\label{TC:ACTC05}}a) infrastructure items that could be misinterpreted as lane markings: curbs \newline
(\TCref{TC:ACTC05}b) infrastructure items that could be misinterpreted as lane markings: dividers 
& 
& 
(further analysis needed) common TCs, conduct simulations to confirm/discard relevance for MPC-TP \\
\hline

(\TCref{TC:AOUCS08}b) irregular road surface: potholes\newline
(\TCref{TC:AOUCS08}c) irregular road surface: road bumps 
& 
& 
(further analysis needed) common TCs, conduct simulations to confirm/discard relevance for MPC-TP \\
\hline

({\TC\label{TC:ACTC09}}a) slippery road surface: wet road with foilage\newline
(\TCref{TC:ACTC09}b) slippery road surface: snowy / icy road
& 
(\FIref{FI:Drivable}a) inability to generate a drivable trajectory in certain situations: due to linear tire model in LDBM
&
(ODD restriction) limit permissible minimum temperature to exclude snowy/icy conditions (\Specref{Spec:ODD}f*)\newline
(refined specification/design) infeasibility analysis: temporarily relax the linear tire regime if the friction value drops by more than $\mu_{\mathrm{d}}$ within $t_{\mathrm{d}}$\newline
(further analysis needed) common TCs, conduct simulations to confirm/discard relevance for MPC-TP \\
\hline

%(\TCref{TC:AOUCS04}a) right-of-way is unclear: at the end of two merging lanes or during special general lane changes
%&
%(\FIref{FI:AOUCS04}a) overly conservative behavior of the MPC-TP: inability to ensure sufficient progress on the road\newline
%(\FIref{FI:AOUCS04}b) overly conservative behavior of the MPC-TP: inability to avoid deadlocks in certain situations\newline
%(\FIref{FI:DangerousManeuver}) generation of trajectories representing unusual, illegal, counter-intuitive, or dangerous maneuvers
%&
%(Interaction-aware MPC) interaction-aware MPC approach, in coordination with the Tactical Planner (\Specref{Spec:InteractionAwareMPC}*)\\
%\\ \hline

\multirow{2}{6cm}{({\TC\label{TC:EAL01}}) entering the freeway from an on-ramp \newline
({\TC\label{TC:EAL02}}) exiting the freeway via an off-ramp \newline
({\TC\label{TC:EAL04}}) lane adding situations}
& 
(\FIref{FI:InsufficientTime}) insufficient time to solve the OCP, in order to compute a new trajectory\newline
(\FIref{FI:Infeasible}) inability to generate a feasible (collision-free) trajectory in certain situations
& 
(refined specification/design) potential extension of computation time to multiple time steps based on the concept of a reaction horizon\newline
(refined specification/design) infeasibility analysis\\
\cline{2-3}
& 
(\FIref{FI:DangerousManeuver}) generation of trajectories representing unusual, illegal, counter-intuitive, or dangerous maneuvers
& 
(further analysis needed) conduct simulations and/or real experiments\\
\hline

\multirow{2}{6cm}{situations with special vehicles or vehicles operating beyond traffic rules:\newline
(\TCref{TC:AOUCS06}) wrong way driver \newline
({\TC\label{TC:EAL05}}) approaching emergency vehicle \newline
({\TC\label{TC:EAL06}}) heavy transport or oversized vehicles \newline
({\TC\label{TC:EAL07}}) vehicle parking on the freeway shoulder\newline 
({\TC\label{TC:EAL08}}) reckless driver / person driving under-the-influence}
& 
(\FIref{FI:DangerousManeuver}) generation of trajectories representing unusual, illegal, counter-intuitive, or dangerous maneuvers
& 
(refined specification/design) introduce specialized driving modes for particular situations (\Specref{Spec:SpecialDrivingModes}*)\newline
(further analysis needed) conduct simulations and/or real experiments\\
\cline{2-3}
& 
(\FIref{FI:InsufficientTime}) insufficient time to solve the OCP, in order to compute a new trajectory\newline
(\FIref{FI:Infeasible}) inability to generate a feasible (collision-free) trajectory in certain situations\newline
& 
(refined specification/design) potential extension of computation time to multiple time steps based on the concept of a reaction horizon\newline
(refined specification/design) infeasibility analysis\\
\hline

\multirow{2}{6cm}{(\TCref{TC:AOUCS08}d) irregular road surface: ruts\newline
(\TCref{TC:AOUCS08}e) irregular road surface: blow-ups}
&
(\FIref{FI:InsufficientTime}) insufficient time to solve the OCP, in order to compute a new trajectory\newline
(\FIref{FI:Infeasible}) inability to generate a feasible (collision-free) trajectory in certain situations
& 
(refined specification/design) potential extension of computation time to multiple time steps based on the concept of a reaction horizon\newline
(refined specification/design) infeasibility analysis
\\
\cline{2-3}
&
(\FIref{FI:DangerousManeuver}) generation of trajectories representing unusual, illegal, counter-intuitive, or dangerous maneuvers
& 
(further analysis) extension of perception required
\\
\hline

({\TC\label{TC:EAL09}}) freeway shoulder not available \newline
({\TC\label{TC:EAL10}}) freeway shoulder to be used as additional lane 
&
(\FIref{FI:DangerousManeuver}) generation of trajectories representing unusual, illegal, counter-intuitive, or dangerous maneuvers
&
(refined specification/design) introduce specialized driving modes for particular situations (\Specref{Spec:SpecialDrivingModes}*)\newline
(further analysis needed) conduct simulations and/or real experiments
\\
\hline

({\TC\label{TC:EAL11}}) end of freeway \newline
 ({\TC\label{TC:EAL12}}) freeway closure
 &
 (\FIref{FI:DangerousManeuver}) generation of trajectories representing unusual, illegal, counter-intuitive, or dangerous maneuvers
&
(further analysis needed) conduct simulations and/or real experiments\\
\hline
 (\TC\label{TC:EAL13}) steep road slope \newline
(\TC\label{TC:EAL14}) high road curvature 
&
(\FIref{FI:Drivable}a) inability to generate a drivable trajectory in certain situations: due to linear tire model in LDBM\newline
(\FIref{FI:AAOD01}) poor approximation of real vehicle behavior by the LDBM in the OCP
&
(further analysis required) experimental analysis: evaluate if MPC-TP behaves safely in these situations; otherwise consider usage of a more complex model (e.g., a two track model) in the OCP\\
\hline

(\TCref{TC:ASA01}) side wind gusts
&
(\FIref{FI:DangerousManeuver}) generation of trajectories representing unusual, illegal, counter-intuitive, or dangerous maneuvers\newline
(\FIref{FI:Infeasible}) inability to generate a feasible (collision-free) trajectory in certain situations\newline
(\FIref{FI:InsufficientTime}) insufficient time to solve the OCP, in order to compute a new trajectory
& 
(no change needed) covered by wind limitation in the ODD (\Specref{Spec:ODD}g)\\
\hline

({\TC\label{TC:EAL15}}) potentially unknown traffic scenarios / traffic signs in other European countries
&
(\FIref{FI:DangerousManeuver}) generation of trajectories representing unusual, illegal, counter-intuitive, or dangerous maneuvers\newline
(\FIref{FI:Infeasible}) inability to generate a feasible (collision-free) trajectory in certain situations\newline
(\FIref{FI:InsufficientTime}) insufficient time to solve the OCP, in order to compute a new trajectory
& 
(no change needed) covered by permissible road type restriction in the ODD (\Specref{Spec:ODD}a) \\
\hline

    \end{longtable}
    \normalsize

%\textcolor{red}{Idea: create a graph that shows the relationships between the identified TCs and FIs.}

\end{landscape}

\newpage
\bibliography{bibcontr,bibeng,bibmath}

\begin{thebibliography}{10}
\expandafter\ifx\csname url\endcsname\relax
  \def\url#1{\texttt{#1}}\fi
\expandafter\ifx\csname urlprefix\endcsname\relax\def\urlprefix{URL }\fi
\expandafter\ifx\csname href\endcsname\relax
  \def\href#1#2{#2} \def\path#1{#1}\fi

\bibitem{MonteEtAl:2006}
M.~Montemerlo, {et al.}, Stanley: The robot that won the {DARPA} {G}rand
  {C}hallenge, Journal of Field Robotics 23~(9) (2006) 661--692.
\newblock \href {http://dx.doi.org/10.1002/rob.20147}
  {\path{doi:10.1002/rob.20147}}.

\bibitem{MonteEtAl:2008}
M.~Montemerlo, {et al.}, Junior: The {S}tanford entry in the urban challenge,
  Journal of Field Robotics 25~(9) (2008) 569--597.

\bibitem{BachaEtAl:2008}
A.~Bacha, C.~Bauman, R.~Faruque, M.~Fleming, C.~Terwelp, C.~Reinholtz, D.~Hong,
  A.~Wicks, T.~Alberi, D.~Anderson, S.~Cacciola, P.~Currier, A.~Dalton,
  J.~Farmer, J.~Hurdus, S.~Kimmel, P.~King, A.~Taylor, D.~V. Covern,
  M.~Webster, Odin: Team {VictorTango}?s entry in the {DARPA} {U}rban
  {C}hallenge, Journal of Field Robotics 25~(8) (2003) 467--492.
\newblock \href {http://dx.doi.org/10.1177/02783649030227008}
  {\path{doi:10.1177/02783649030227008}}.

\bibitem{PadenEtAl:2016}
B.~Paden, M.~{\v C}{\'a}p, S.~Z. Yong, D.~Yershov, E.~Frazzoli, A survey of
  motion planning and control techniques for self-driving urban vehicles, IEEE
  Transactions on Intelligent Vehicles 1 (2016) 33--55.
\newblock \href {http://dx.doi.org/10.1109/TIV.2016.2578706}
  {\path{doi:10.1109/TIV.2016.2578706}}.

\bibitem{DixitEtAl:2020}
S.~Dixit, U.~Montanaro, M.~Dianati, D.~Oxtoby, T.~Mizutani, A.~Mouzakitis,
  S.~Fallah, Trajectory planning for autonomous high-speed overtaking in
  structured environments using {Robust} {MPC}, IEEE Transactions on
  Intelligent Transportation Systems 21~(6) (2020) 2310--2323.
\newblock \href {http://dx.doi.org/10.1109/TITS.2019.2916354}
  {\path{doi:10.1109/TITS.2019.2916354}}.

\bibitem{MusaEtAl:2021}
A.~Musa, M.~Pipicelli, M.~Spano, F.~Tufano, F.~{De Nola}, G.~{Di Blasio},
  A.~Gimelli, D.~A. Misul, G.~Toscano, A review of {M}odel {P}redictive
  {C}ontrols applied to advanced driver-assistance systems, Energies 14~(23)
  (2021) 7974--7997.
\newblock \href {http://dx.doi.org/10.3390/en14237974}
  {\path{doi:10.3390/en14237974}}.

\bibitem{Mayne:2000}
D.~Q. Mayne, J.~B. Rawlings, C.~Rao, P.~O. Scokaert, Constrained model
  predictive control: Stability and optimality, Automatica 36~(6) (2000)
  789--814.

\bibitem{GruenePannek:2011}
L.~Gr{\"u}ne, J.~Pannek, Nonlinear Model Predictive Control, Springer, London
  et al., 2011.

\bibitem{BoBeMo:2017}
F.~Borrelli, A.~Bemporad, M.~Morari, Predictive Control for Linear and Hybrid
  Systems, Cambridge University Press, Cambridge, United Kingdom, 2017.
\newblock \href {http://dx.doi.org/10.1017/978-1-139-06175-9}
  {\path{doi:10.1017/978-1-139-06175-9}}.

\bibitem{RaMaDi:2018}
J.~B. Rawlings, D.~Q. Mayne, M.~M. Diehl, Model Predictive Control: Theory,
  Computation, and Design, 2nd Edition, Nob Hill Publishing, Madison (WI),
  United States, 2018.

\bibitem{GoetteEtAl:2016}
C.~G{\"o}tte, M.~Keller, C.~Ha{\ss}, T.~Bertram, Model predictive planning and
  control applied to critical traffic situations, ATZ Worldwide 118~(9) (2016)
  64--69.

\bibitem{FranzeLucia:2016}
G.~Franz{\`e}, W.~Lucia, A receding horizon control strategy for autonomous
  vehicles in dynamic environments, IEEE Transactions on Control Systems
  Technology 24 (2016) 695--702.
\newblock \href {http://dx.doi.org/10.1109/TCST.2015.2440999}
  {\path{doi:10.1109/TCST.2015.2440999}}.

\bibitem{FraschEtAl:2013}
J.~V. Frasch, A.~Gray, M.~Zanon, H.-J. Ferreau, S.~Sager, F.~Borrelli,
  M.~Diehl, An auto-generated nonlinear {MPC} algorithm for real-time obstacle
  avoidance of ground vehicles, in: European Control Conference, Zurich,
  Switzerland, 2013, pp. 4136--4141.

\bibitem{KongEtAl:2015}
J.~Kong, M.~Pfeiffer, G.~Schildbach, F.~Borrelli, Kinematic and dynamic vehicle
  models for autonomous driving control design, in: IEEE Intelligent Vehicles
  Symposium, Seoul, Korea, 2015, pp. 1094--1099.

\bibitem{GutjahrEtAl:2017}
B.~Gutjahr, L.~Gr{\"o}ll, M.~Werling, Lateral vehicle trajectory optimization
  using constrained linear time-varying {MPC}, IEEE Transactions on Intelligent
  Transportation Systems 18~(6) (2017) 1586--1595.
\newblock \href {http://dx.doi.org/10.1109/TITS.2016.2614705}
  {\path{doi:10.1109/TITS.2016.2614705}}.

\bibitem{Carvalho:2016}
A.~Carvalho, Predictive control under uncertainty for safe autonomous driving:
  Integrating data-driven forecasts with control design, {Ph.D.}~dissertation,
  University of California Berkeley, Berkeley (CA), United States (2016).

\bibitem{Schildi:2015}
G.~Schildbach, F.~Borrelli, Scenario model predictive control for lane change
  assistance on highways, in: IEEE Intelligent Vehicles Symposium, Seoul, South
  Korea, 2015, pp. 611--616.

\bibitem{CesariEtAl:2017}
G.~Cesari, A.~Carvalho, G.~Schildbach, F.~Borrelli, Scenario model predictive
  control for lane change assistance and autonomous driving on highways, IEEE
  Intelligent Transportation Systems Magazine 9~(3) (2017) 23--35.

\bibitem{ProDriver}
S.~Longo, Virtual driver for autonomous vehicles,
  \url{https://www.embotech.com/products/prodriver/overview}, accessed:
  2023-08-06.

\bibitem{CarvalhoEtAl:2015}
A.~Carvalho, S.~Lef{\`e}vre, G.~Schildbach, J.~Kong, F.~Borrelli, Automated
  driving, the role of forecasts and uncertainty -- a control perspective,
  European Journal of Control 24 (2015) 14--32.

\bibitem{LiuEtAl:2022}
K.~Liu, N.~Li, H.~E. Tseng, I.~Kolmanovsky, A.~Girard, Interaction-aware
  trajectory prediction and planning for autonomous vehicles in forced merge
  scenarios, IEEE Transactions on Intelligent Transportation Systems 24~(1)
  (2023) 474--488.
\newblock \href {http://dx.doi.org/10.1109/TITS.2022.3216792}
  {\path{doi:10.1109/TITS.2022.3216792}}.

\bibitem{GuptaEtAl:2023}
P.~Gupta, D.~Isele, D.~Lee, S.~Bae, Interaction-aware trajectory planning for
  autonomous vehicles with analytic integration of {N}eural {N}etworks into
  {M}odel {P}redictive {C}ontrol, arXiv Preprint (2023) 1--7\href
  {http://dx.doi.org/10.48550/arXiv.2301.05393}
  {\path{doi:10.48550/arXiv.2301.05393}}.

\bibitem{NairEtAl:2022}
S.~H. Nair, V.~Govindarajan, T.~Lin, Y.~Wang, E.~H. Tseng, F.~Borrelli,
  Stochastic {MPC} with dual control for autonomous driving with multi-modal
  interaction-aware predictions, in: 15th International Symposium on Advanced
  Vehicle Control, Kanagawa, Japan, 2022.

\bibitem{BencioliniEtAl:2023}
T.~Benciolini, D.~Wollherr, M.~Leibold, Non-conservative trajectory planning
  for automated vehicles by estimating intentions of dynamic obstacles, IEEE
  Transactions on Intelligent Vehicles 8~(3) (2023) 2463--2481.
\newblock \href {http://dx.doi.org/10.1109/TIV.2023.3234163}
  {\path{doi:10.1109/TIV.2023.3234163}}.

\bibitem{ChenEtAl:2022}
Y.~Chen, U.~Rosolia, W.~Ubellacker, N.~Csomay-Shanklin, {A}aron D.~{A}mes,
  Interactive multi-modal motion planning with {B}ranch {M}odel {P}redictive
  {C}ontrol, IEEE Robotics and Automation Letters 7~(2) (2022) 5365--5372.
\newblock \href {http://dx.doi.org/10.1109/LRA.2022.3156648}
  {\path{doi:10.1109/LRA.2022.3156648}}.

\bibitem{OliveiraEtAl:2023}
R.~Oliveira, S.~H. Nair, B.~Wahlberg, Interaction and decision making-aware
  motion planning using {B}ranch {M}odel {P}redictive {C}ontrol, arXiv Preprint
  (2023) 1--8\href {http://dx.doi.org/10.48550/arXiv.2302.00060}
  {\path{doi:10.48550/arXiv.2302.00060}}.

\bibitem{ISO21448:2022}
{ISO 21448:2022 Road vehicles -- Safety of the intended functionality},
  International standard, International Organization for Standardization,
  Geneva, CH (Jun. 2022).

\bibitem{9724010}
R.~Zhu, A.~Gu, Z.~Wu, B.~Liu, M.~Yu, Research on sotif of automatic driving
  system, in: 2022 14th International Conference on Measuring Technology and
  Mechatronics Automation (ICMTMA), 2022, pp. 228--231.
\newblock \href {http://dx.doi.org/10.1109/ICMTMA54903.2022.00051}
  {\path{doi:10.1109/ICMTMA54903.2022.00051}}.

\bibitem{9872207}
Z.~Qidong, Z.~Tong, Z.~Yunshuang, C.~Chao, Z.~Qingyu, Z.~Shuai, D.~Zhibin, The
  research on the identification of acc sotif triggering conditions based on
  scenario analysis, in: 2022 IEEE International Conference on Real-time
  Computing and Robotics (RCAR), 2022, pp. 263--266.
\newblock \href {http://dx.doi.org/10.1109/RCAR54675.2022.9872207}
  {\path{doi:10.1109/RCAR54675.2022.9872207}}.

\bibitem{9872242}
L.~Junfeng, Z.~Yunshuang, Z.~Shuai, C.~Chao, D.~Zhibin, A research on sotif of
  lka based on stpa, in: 2022 IEEE International Conference on Real-time
  Computing and Robotics (RCAR), 2022, pp. 396--400.
\newblock \href {http://dx.doi.org/10.1109/RCAR54675.2022.9872242}
  {\path{doi:10.1109/RCAR54675.2022.9872242}}.

\bibitem{BaerEtAl:2009}
M.~Baer, M.~E. Bouzouraa, C.~Demiral, U.~Hofmann, S.~Gies, K.~Diepold,
  Egomaster: A central ego motion estimation for driver assist systems, in:
  IEEE International Conference on Control and Automation, Christchurch, New
  Zealand, 2009, pp. 1708--1715.
\newblock \href {http://dx.doi.org/10.1109/ICCA.2009.5410518}
  {\path{doi:10.1109/ICCA.2009.5410518}}.

\bibitem{ThornEtAl2018}
E.~Thorn, M.~Kimmel, Shawn C.and~Chaka, A framework for automated driving
  system testable cases and scenarios (rep. no dot hs 812 623), Tech. rep.,
  National Highway Traffic Safety Administration, Washington, DC (2018).

\bibitem{ISO34503:2023}
{ISO 34503: Road Vehicles -- Test scenarios for automated driving systems --
  Specification for operational design domain}, International standard,
  {International Organization for Standardization}, Geneva, Switzerland (2023).

\bibitem{PEGASUS:2019}
{PEGASUS} method -- an overview, Project report, German Aerospace Center,
  Braunschweig, Germany (2019).

\bibitem{LiuLeeEtAl:2017}
C.~Liu, S.~Lee, S.~Varnhagen, H.~E. Tseng, Path planning for autonomous
  vehicles using {M}odel {P}redictive {C}ontrol, in: IEEE Intelligent Vehicles
  Symposium, Redondo Beach (CA), United States, 2017, pp. 174--179.
\newblock \href {http://dx.doi.org/10.1109/IVS.2017.7995716}
  {\path{doi:10.1109/IVS.2017.7995716}}.

\bibitem{YuEtAl:2021}
S.~Yu, M.~Hirche, Y.~Huang, H.~Chen, F.~Allg{\"o}wer, {M}odel {P}redictive
  {C}ontrol for autonomous ground vehicles: A review, Autonomous Intelligent
  Systems 1~(4) (2021) 1--17.
\newblock \href {http://dx.doi.org/10.1007/s43684-021-00005-z}
  {\path{doi:10.1007/s43684-021-00005-z}}.

\bibitem{MicheliEtAl:2022}
F.~Micheli, M.~Bersani, S.~Arrigoni, F.~Braghin, F.~Cheli, {NMPC} trajectory
  planner for urban autonomous driving, Vehicle System Dynamics 0~(0) (2022)
  1--23.
\newblock \href {http://dx.doi.org/10.1080/00423114.2022.2081220}
  {\path{doi:10.1080/00423114.2022.2081220}}.

\bibitem{Waechter:2010}
A.~W{\"a}chter, Introduction to {IPOPT}: A tutorial for downloading,
  installing, and using {IPOPT}, Technical report, IBM T.J.\ Watson Research
  Center, Yorktown Heights (NY), United States, available at
  \texttt{http://web.mit.edu/ipopt\_v3.8/doc/documentation.pdf} (2010).

\bibitem{Schildi:2016}
G.~Schildbach, A new nonlinear model predictive control algorithm for vehicle
  path tracking, in: Symposium on Advanced Vehicle Control, Munich, Germany,
  2016, pp. 139--144.

\bibitem{GraichenEtAl:2019}
T.~Englert, A.~V{\"o}lz, F.~Mesmer, S.~Rhein, K.~Graichen, A software framework
  for embedded {N}onlinear {M}odel {P}redictive {C}ontrol using a
  gradient-based augmented {L}agrangian approach ({GRAMPC}), Optimization and
  Engineering 20 (2019) 769--809.
\newblock \href {http://dx.doi.org/10.1007/s11081-018-9417-2}
  {\path{doi:10.1007/s11081-018-9417-2}}.

\bibitem{ZanelliEtAl:2020}
A.~Zanelli, A.~Domahidi, J.~L. Jerez, M.~Morari, {FORCES NLP}: An efficient
  implementation of interior-point methods for multistage nonlinear nonconvex
  programs, International Journal of Control 93~(1) (2020) 13--29.
\newblock \href {http://dx.doi.org/10.1080/00207179.2017.1316017}
  {\path{doi:10.1080/00207179.2017.1316017}}.

\bibitem{ISO26262:2018}
{ISO 26262:2018 Road vehicles -- Functional safety}, International standard,
  International Organization for Standardization, Geneva, CH (Dec. 2018).

\bibitem{Con18}
M.~Conrad, \href{https://vipread.com/library/topic/2054}{Automated driving:
  Challenges in the interplay between functional safety and safety of the
  intended functionality}, presented at SAE Automated Vehicle Security and
  Safety Technology Forum (SAE-AWC 2018), Shenzen, CH (2018).
\newline\urlprefix\url{https://vipread.com/library/topic/2054}

\bibitem{Werling:2011}
M.~Werling, Ein neues {K}onzept f{\"u}r die {T}rajektoriengenerierung und
  -stabilisierung in zeitkritische {V}erkehrsszenarien, {Ph.D.}~dissertation,
  Karlsruher Institut f{\"u}r Technologie, Karlsruhe, Germany (2011).

\bibitem{LubinieckiEtAl:2020}
T.~Lubiniecki, S.~Beer, A.~Meisinger, F.~Sellmann, P.~Spannaus, G.~Schildbach,
  Concept and implementation of an optimization-based safety verification
  strategy for a trajectory following controller, in: T.~Bertram (Ed.),
  Automatisiertes Fahren 2020, Springer Vieweg, Wiesbaden, Germany, 2021, pp.
  111--124.
\newblock \href {http://dx.doi.org/10.1007/978-3-658-34752-9_10}
  {\path{doi:10.1007/978-3-658-34752-9_10}}.

\bibitem{SwaHed:1996}
D.~Swaroop, J.~K. Hedrick, String stability of interconnected systems, IEEE
  Transactions on Automatic Control 41~(3) (1996) 349--357.
\newblock \href {http://dx.doi.org/10.1109/9.486636}
  {\path{doi:10.1109/9.486636}}.

\bibitem{ShawHed:2007}
E.~Shaw, J.~K. Hedrick, String stability analysis for heterogeneous vehicle
  strings, in: American Control Conference, New York (NY), United States, 2007,
  pp. 3118--3125.
\newblock \href {http://dx.doi.org/10.1109/ACC.2007.4282789}
  {\path{doi:10.1109/ACC.2007.4282789}}.

\bibitem{WolfeEtAl:2020}
B.~Wolfe, B.~Seppelt, B.~Mehler, B.~Reimer, R.~Rosenholtz, Rapid holistic
  perception and evasion of road hazards, Journal of Experimental Psychology:
  General 149~(3) (2020) 490--500.
\newblock \href {http://dx.doi.org/10.1037/xge0000665}
  {\path{doi:10.1037/xge0000665}}.

\end{thebibliography}

\end{document}